\documentclass[twocolumn,floatfix,superscriptaddress]{revtex4-2}
\setcitestyle{super}

\usepackage{xfrac}
\usepackage{bm}

\usepackage{upgreek}

\usepackage{lettrine}

\usepackage[normalem]{ulem}

\usepackage{fancyhdr}
\usepackage{amsmath}
\usepackage{graphicx}
\usepackage{color}

\usepackage{lettrine}

\renewcommand{\figurename}{\textbf{Fig.}}
\renewcommand\thefigure{\textbf{\arabic{figure}}}

\usepackage[unicode=true,bookmarks=true,pdfborder={0 0 0},backref=true,colorlinks=true,allcolors=blue,anchorcolor=black]{hyperref}
\usepackage{breakurl}

\makeatletter

\@ifundefined{textcolor}{}
{
 \definecolor{BLACK}{gray}{0}
 \definecolor{WHITE}{gray}{1}
 \definecolor{RED}{rgb}{0.7,0,0}
 \definecolor{ORANGE}{rgb}{1,0.25,0}
 \definecolor{GREEN}{rgb}{0,1,0}
 \definecolor{BLUE}{rgb}{0,0,1}
 \definecolor{CYAN}{cmyk}{1,0,0,0}
 \definecolor{MAGENTA}{cmyk}{0,1,0,0}
 \definecolor{YELLOW}{cmyk}{0,0,1,0}
}

\newcommand{\nanotec}{
CNR NANOTEC, Istituto di Nanotecnologia, Via Monteroni, 73100 Lecce, Italy
}

\newcommand{\warsav}{
Institute of Physics, Polish Academy of Sciences, Al. Lotnik\'{o}w 32/46, 02-668 Warsaw, Poland
}

\newcommand{\marseille}{
Aix Marseille Universit\'{e}, CNRS, Centrale Marseille, LMA UMR 7031 Marseille, France
}

\newcommand{\unisalento}{
Dipartimento di Matematica e Fisica E. De Giorgi, Università del Salento, Campus Ecotekne, via Monteroni, 73100 Lecce, Italy
} 

\newcommand{\wolver}{
Faculty of Science and Engineering, University of Wolverhampton, Wulfruna Street, WV1 1LY Wolverhampton, UK
}

\newcommand{\rqc}{
Russian Quantum Center, Skolkovo innovation city, 121205 Moscow, Russia
}

\newcommand{\moscow}{
National Research Nuclear University MEPhI (Moscow Engineering Physics Institute), 115409 Moscow, Russia\\
*email \href{mailto:lorenzo.dominici@nanotec.cnr.it}{lorenzo.dominici@nanotec.cnr.it},
\href{mailto:nsvoronova@mephi.ru}{nsvoronova@mephi.ru}.
Cite this work as \href{https://www.nature.com/articles/s42005-023-01305-x}{Commun.~Phys.~6, 197 (2023)}.
}

\renewcommand\frontmatter@abstractwidth{\dimexpr\textwidth-0.5in\relax}
\makeatother

\begin{document}

\title{
\usefont{OT1}{cmss}{m}{n}
{\Large
\textbf{Coupled quantum vortex kinematics and Berry curvature in real space
}
}
}

\author{Lorenzo Dominici*}
\affiliation{\nanotec}

\author{Amir Rahmani}
\affiliation{\warsav}

\author{David Colas}
\affiliation{\marseille}



\author{Dario Ballarini}
\affiliation{\nanotec}

\author{Milena De Giorgi}
\affiliation{\nanotec}

\author{Giuseppe Gigli}
\affiliation{\nanotec}
\affiliation{\unisalento}

\author{Daniele Sanvitto}
\affiliation{\nanotec}

\author{Fabrice P.~Laussy}
\affiliation{\wolver}
\affiliation{\rqc}

\author{Nina Voronova*,}
\affiliation{\rqc}
\affiliation{\moscow}

\begin{abstract}
\textbf{Abstract.} 
The Berry curvature provides a powerful tool to unify several branches
of science through their geometrical aspect: topology, energy bands,
spin and vector fields. While quantum defects--phase vortices and
skyrmions--have been in the spotlight, as rotational entities in
condensates, superfluids and optics, their dynamics in multi-component
fields remain little explored. Here we use two-component microcavity
polaritons to imprint a dynamical pseudospin texture in the form of a double full Bloch beam, 
a conformal continuous vortex beyond unitary skyrmions. 
The Berry curvature plays a key role to link various
quantum spaces available to describe such textures. It explains for
instance the ultrafast spiraling in real space of two singular vortex
cores, providing in particular a simple expression--also involving the
complex Rabi frequency--for their intricate velocity. Such Berry
connections open new perspectives for understanding and controlling 
highly-structured quantum objects, including strongly asymmetric cases 
or even higher multi-component fields.
\end{abstract}


\maketitle

\usefont{OT1}{ppl}{m}{n}






\noindent \textbf{INTRODUCTION}\\
Quantized vortices embody the essence of circular motion in all 
fields and fluids that can be described by an oscillating wavefunction, such as superfluids, superconductors, atomic Bose-Einstein condensates (BECs) and optical fields~\cite{salomaa_quantized_1987,vollhardt_superfluid_1990,lounasmaa_vortices_1999,
volovik_universe_2009,leggett_superfluidity_1999,aranson_world_2002,simula_quantised_2019,shen_optical_2019}.
The complex-valued nature of the wavefunction implies an integer number of phase windings, or wave crests, around the core of such a vortex, contributing to the intrinsic orbital angular momentum (OAM)~\cite{allen_orbital_1992,soskin_topological_1997,milonni_momentum_2010}. 
Unlike classical rigid rotations, the rotating quantum fluid takes on greater momentum and velocity closer to the core of rotation~\cite{pitaevskii_bose-einstein_2016}, as in a gravitational orbital motion. In an optical vortex this applies to the group velocity, differently from the phase velocity. An isolated Bose-Einstein condensate rotating around a single or multiply charged central vortex can be thought of as a fundamental gyroscope~\cite{hodby_experimental_2003}, candidating them as sensitive detectors for gravitational waves when implemented in an orbiting laboratory~\cite{aveline_observation_2020}. At different levels, analogies were drawn, both in the past and recently, to point-like atoms~\cite{thomson_4_1869,karch_particle-vortex_2016} or other elementary particles dressed with two quantum numbers and capable of tunable pair-wise interactions~\cite{zhao_pattern_2017,dominici_interactions_2018} in the nonlinear regime. Quantum vortices have been mostly studied when dealing with nonlinear fluids and their phase transitions, relevant in their observed macroscopic degree of coherence~\cite{dagvadorj_nonequilibrium_2015}, in quasi-two-dimensional (2D) as well as in three-dimensional (3D) fields, giving rise to vortex tubes, networks, rings and 
knots~\cite{leach_vortex_2005,dennis_isolated_2010}. These entities are 
called wave dislocations~\cite{nye_dislocations_1974} or topological defects, because the phase singularity at their core and the order parameter winding around it are features of all waves~\cite{berry_making_2000} at the basis of many structured objects of nontrivial geometry. When implemented in photonics, 
vortices 
represent
a further degree of freedom~\cite{molina-terriza_twisted_2007,yao_orbital_2011}, 
for data multiplexing and for free-air transmission~\cite{krenn_twisted_2016,yan_high-capacity_2014,gibson_free-space_2004} as well as 
for many optical tweezers applications~\cite{padgett_tweezers_2011,he_direct_1995} and structured light schemes~\cite{forbes_structured_2021,secor_complex_2016}. Further dressed with the spin angular momentum (SAM) or polarization degree of freedom, they can create complex textures 
around their core~\cite{lopez-mago_overall_2019,gobel_beyond_2021}, 
subtending 2D skyrmion (baby-skyrmion) and even 3D particle-like optical skyrmions~\cite{sugic_particle-like_2021,parmee_optical_2022,shen_topological_2023} eventually evolving in time~\cite{donati_twist_2016} thanks to different interactions with a material medium. These objects can be of different type, such as Bloch-, Neel- and anti-skyrmions, or strained into a combination of multiple subparticles known as merons~\cite{gobel_beyond_2021}. Recently, 
there is a rising interest in the generation of optical skyrmion beams~\cite{sugic_particle-like_2021,parmee_optical_2022,shen_topological_2023,donati_twist_2016,gutierrez-cuevas_optical_2021,shen_generation_2022,lin_microcavity-based_2021,liu_disorder-induced_2022},
thanks to their tunability and possible use in data encoding and transmission.
This interest is further motivated by the fact that,
both in condensed matter and in optics, even the most fundamental skyrmion in the linear regime can be
viewed as a full Poincar\'{e}~\cite{beckley_full_2010} or full Bloch beam~\cite{dominici_full-bloch_2021}, highlighting their topological connections.

Full Poincaré or Bloch beams can be created using a combination of vortical and vortex-free excitation beams, whose wavefunction simultaneously comprises all the possible quantum states of the associated Hilbert space~\cite{dominici_full-bloch_2021}: a sphere of polarizations or pseudospins. %
They belong to the class of continuous vortices, also known as filled-core, bright core, coreless or non-singular vortices in condensed matter and half-vortices in optics and polaritons \cite{Rubo2007}. The Mermin-Ho~\cite{mermin_circulation_1976} and Anderson-Toulouse vortices~\cite{anderson_phase_1977} in superfluid $^3\text{He-A}$ are also examples of continuous vortices representing half- and full-texture states, respectively. In that case, the pseudospin is the local unit vector of relative orbital angular momentum of a Cooper pair. 
In a ferromagnetic BEC, similar structures may exist~\cite{mizushima_mermin-ho_2002}, where the texture refers to the unit vector parallel to the local magnetization~\cite{yukawa_su_2023}. In these two examples, the pseudospin vector represents a direction in real space.
In general, in any two-state system, such as, for example, two-component atomic BECs or polarized optical fields, 
the associated 
Hilbert space is a four-dimensional space~\cite{dandoloff_two-level_1992}.
Apart from the degrees of freedom represented by the total density 
and the total phase, 
it can be interesting 
to focus on the remaning spinor order parameter (two-dimensional Hilbert space of pseudospins). 
This expresses the relative amplitude and phase between the two system components, that can be associated to the polar $\theta$ and azimuthal $\varphi$ angles on a sphere, \textit{i.e.}, in terms of Stokes-like parameters. 
Their quantum geometry can be conveniently described 
by mapping the sphere texture onto a given parameter space, such as the real or momentum space.
The density of the mapped texture is represented by the so-called Mermin-Ho texture~\cite{mermin_circulation_1976,mermin_topological_1979} or Berry curvature~\cite{Berry_1989}, 
a quantity that, being integrated over the whole target space, 
yields the net number of topological 
wraps around the sphere, or skyrmion number~\cite{xiao,everschor-sitte_real-space_2014,parmee_optical_2022}.
Compared to full Poincar\'{e} beam,
the full Bloch beam changes dynamically,
because its texture is composed of two eigenmodes with different energies.
This results in the coherent oscillations of the two components 
(\textit{e.g.}, the strongly coupled exciton and photon fields of microcavity polaritons),
and can be naturally associated with counter-intuitive physical concepts such as time-varying OAM,
a feature that has recently attracted attention~\cite{rego_generation_2019,dominici_full-bloch_2021}.  
Indeed, 
the coherent exchange of energy and momentum between the components 
also manifests in a dynamical offset of the vortex cores from the centre in these composite beams.
This provides a nice visualisation of what happens to 
the wavefunction in the case of time-varying OAM,
implementing the fact that the mean OAM per particle is not directly determined by the presence of a given number of vortices, but depends on their positions, their motion and the surrounding density~\cite{chevy_measurement_2000,madison_vortex_2000,berry_paraxial_1998,
Satyajit_geometric_2019,hosseini_temporal_2020,choi_observation_2022}.
This could be used for the direct encoding of specific OAM waveforms.

Here, we study a 
double full Bloch beam, 
created 
in a polariton fluid,
and track its evolution that takes the form of spiraling vortices, driven by a mechanism which we clarify in the following.
Microcavity polaritons are hybrid 
light-matter quasiparticles arising from the strong coupling of microcavity (MC) photons and quantum well (QW) excitons. This makes them a two-component system (even in the case of homogeneous polarization). 
Differently from a polarization pseudospin, the polariton pseudospin represents the composition of the full wavefunction in terms of the two normal modes of the system, 
upper and lower polaritons (UP and LP).
They inherit nonlinearity from the excitons, interacting composite bosons~\cite{combescot_excitons_2015}, but show interesting aspects also in the linear regime. Polaritons provide a powerful testbed for many quantum-fluids phenomena~\cite{Byrnes2014} spanning from Bose-Einstein condensation, long range order coherence, phase transitions~\cite{dagvadorj_nonequilibrium_2015} and superfluidity, as well as 
spontaneous or imprinted 
vortex structures formation and dynamics~\cite{dominici_vortex_2015}. Their open-dissipative nature allows for both the resonant excitation by means of continuous-wave or pulsed laser light, as well as for detection of their state via the emitted light. 

In the present setting,
the double vortex is imprinted by a photon pulse and perturbed by a second pulse which is vortex-free. This splits the two initally singular vortex into four coreless vortices (in the polariton pseudospin texture), representing the initial state. Its evolution is imaged by means of a digital off-axis holography scheme~\cite{dominici_ultrafast_2014} allowing to retrieve both the amplitude and phase maps of the emission in ultrafast time. 
The observed dynamics happens in the linear regime and is enabled thanks to the natural assets of polaritons, that is their Rabi oscillations~\cite{dominici_ultrafast_2014} and radiative decay.
In fact, the motion of photonic vortices 
inside the composite vortex~\cite{maleev_composite_2003}
is due to the spatially varying relative phase and amplitude 
between the UP and LP mode,
and their drift in time due to the different eigenfrequencies. 
This is possible upon associating the photonic vortex core 
to a specific pseudospin on the equator of the Bloch sphere of polariton states,
and mapping the pseudospins by means of the Bloch sphere metric in real space.
In the case of a unitary full Bloch beam~\cite{dominici_full-bloch_2021} (when the excitation pulse carries a unitary vortex charge), 
the mapping is a conformal stereographic projection, and the dynamics consists of reshaping Apollonian circles. In the case of a moving packet~\cite{dominici_shaping_2021}  (when the excitation imprints a moving vortex upon oblique incidence), the mapping is neither conformal nor stereographic, but involves vortex pair creation and annihilation events.
Here, 
we study the dynamical texture of a \textit{double} full Bloch beam of polaritons, and show that it is conformal despite not being a stereographic projection. Consequently, the two spiraling vortex cores are following reshaping orbits that are not circular. This complicated motion can be described in terms of the two orthogonal velocities on the sphere and in real space. We use this to reveal a general property for the superposition of Laguerre-Gaussian beams (LGs). 

Furthermore, 
our study is also the first general description, to the best of our knowledge, of the real-space Berry curvature of a polariton state, 
associated to their Bloch sphere pseudospin mapping. 
Its integral allows us to show that the topological charge of the dynamical double full Bloch beam is indeed two and that it is conserved in time, despite the metric spinning and the Berry curvature itself constantly reshaping. 
Conceptually, this brings the notion of a pseudospin texture manifesting as an observable point-like object (the photonic vortex core) with a conserved topological number.
Finally, we derive a direct connection between the rate of change of the polariton state on the Bloch sphere and the velocity of the pseudospins in real space. This further simplifies into an elegant and intriguing expression for the instantaneous speed of the observable vortex cores in real space, 
only depending on the complex Rabi frequency and the 
local Berry curvature. Such a link is inveresely proportional to the curvature, differently from other more known cases.\\ 

\begin{figure}[t!]
  \centering \includegraphics[width=1.05\linewidth]{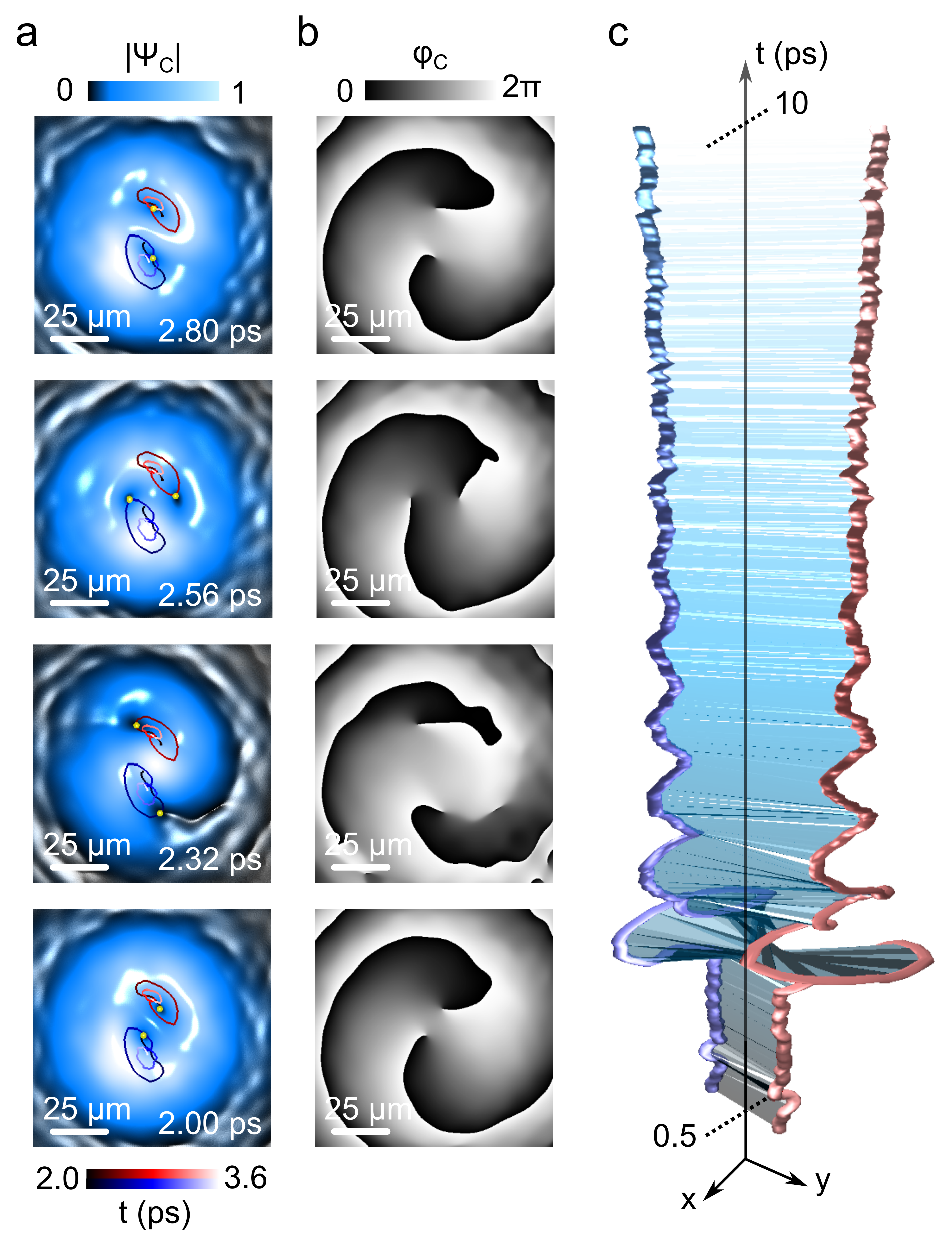}
  \caption{\textbf{Double Rabi spiraling vortex.}
    \textbf{a},\textbf{b}, Experimental amplitude and phase maps of the photonic emission from the polariton fluid, when exciting with a double vortex ($l_A = 2$) and after arrival of a plain Gaussian pulse ($l_B = 0$) at the time $t_{AB}=2.04$~ps. The different frames correspond to times $t = 2.0, 2.32, 2.56$ and $2.80~\text{ps}$. The instantaneous phase singularities are tracked as yellow dots in the amplitude maps, and the solid lines represent the vortex trajectories drawn 
    over two Rabi-oscillations periods ($t = 2.0-3.6~\text{ps}$, red and blue lines are used 
    to distinguish the two 
    different cores in the photon emission).
    \textbf{c}, Vortex lines plotted as $xyt$ curves (time range $t=0.5-10\text{ ps}$, step $\delta t=0.02\text{ ps}$). See also the Supplementary Movie 1.
    }
\label{FIG_exp_time}
\end{figure}

\noindent \textbf{RESULTS AND DISCUSSION}\\
\noindent \textbf{Spiraling of the polariton double vortex}\\
By means of a $q$-plate carrying a double topological charge, we shape the photonic
pulse $A$ into a modified Laguerre-Gauss $\text{LG}_{02}$ state~\cite{dominici_interactions_2018}. Its double winding is translated into two unitary co-winding vortices, whose initial separation can be controlled by the tuning of the $q$-plate device. We exploit this effect to realize a polariton fluid with two spatially separated phase singularities, introducing the potentiality for an asymmetry factor with respect to the following dynamics.
After the first pulse, the vortex cores are stably located in the inner region of the spot, split by about 12~$\mu$m along an oblique direction (\textit{i.e.}, they are off-axis with respect to the center of the beam), as shown in the bottom panels of Fig.~\ref{FIG_exp_time}a,b reporting the amplitude (a) and phase (b) of the double vortex state at the time $t = 2.0$~ps, before the arrival of the 
second pulse $B$, which is a plain Gaussian, $\text{LG}_{00}$. Upon the overlap with the pulse $B$, the vortices are 
displaced even more off-axis in opposite directions and reach a maximum distance of $\approx 40~\mu$m (at $t = 2.32$~ps), starting a rotational motion. 
We note that similarly to a precession, while the vortices move on the ps time scale, the field is rotating around each core at the optical frequency in the fs time scale.
They reach a horizontal alignment ($t = 2.56$~ps) before coming back to the central region after one Rabi-oscillation period, which is shown in the top panels ($t = 2.80$~ps). Their trajectories, extended into the next Rabi period ($t = 2.0$ -- 3.6~ps), are reported as superimposed solid lines (red and blue). In summary, the two $xyt$ vortex lines start a damped spiraling motion to end up at two new positions, and this can be clearly seen when reporting them as a 3D perspective in Fig.~\ref{FIG_exp_time}c. 

\begin{figure*}[htbp]
  \centering \includegraphics[width=0.75\linewidth]{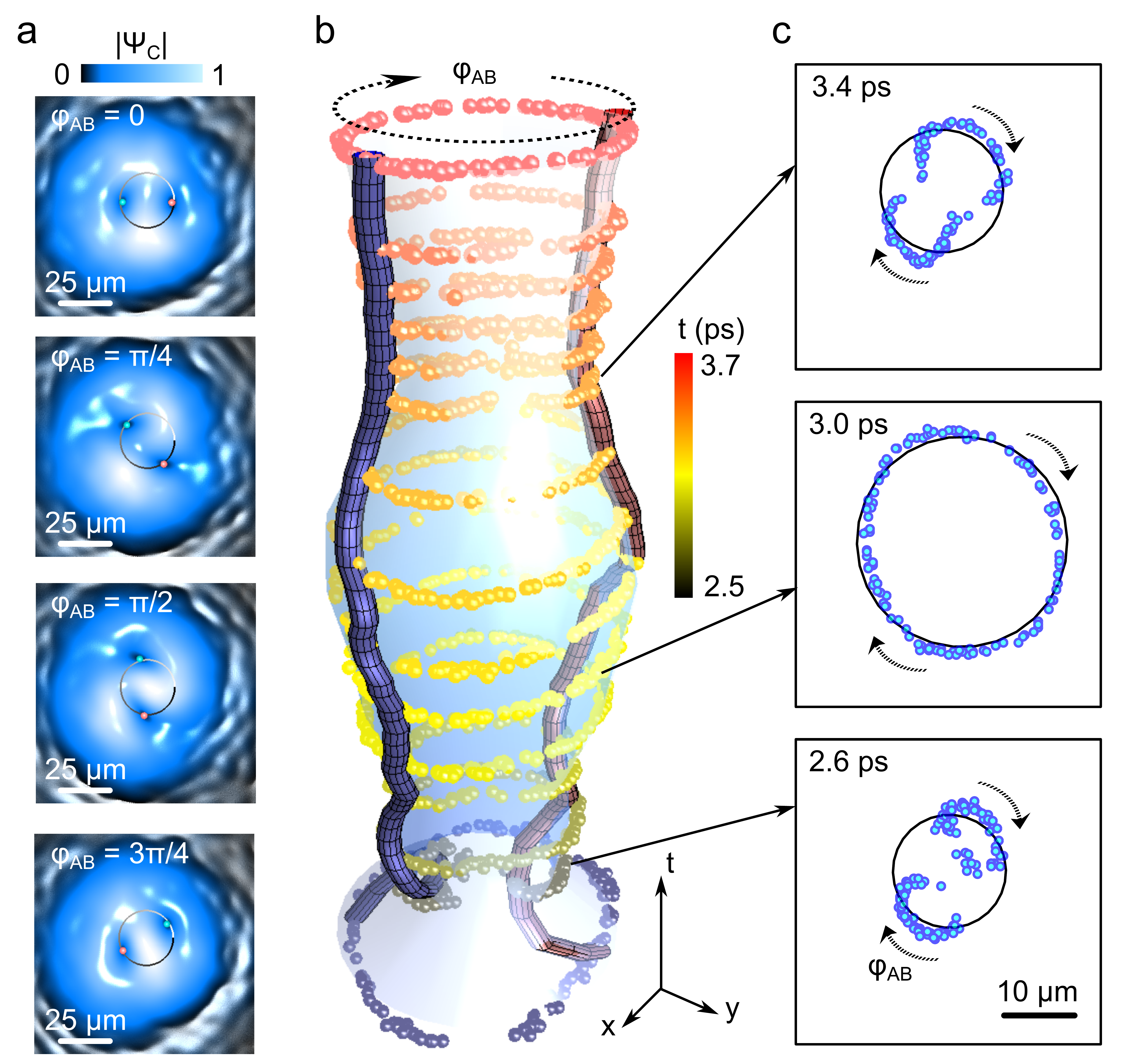}
  \caption{\textbf{Phase delay control of vortex lines.}
    \textbf{a}, Polariton amplitude maps in the phase delay experiment. The snapshots are taken at $t = 3.7$~ps, with four different phase delays $\varphi_{AB}$ spaced by $\uppi/4$. The phase singularities have been marked with blue/red dots in the amplitude maps, and the black/white line is the circle fitting their positions.
    \textbf{b}, The specific $xyt$ vortex lines (at a given phase delay) of the two unitary vortices (blue and red tubes) during one and half Rabi-oscillations period (time range $t = 2.5 - 3.7$~ps). The topological bottle surface described by the vortex cores when sweeping the optical phase delay between the two pulses has been mapped (by spheres) at 100~fs or 50~fs time intervals, spanning $\varphi_{AB}$ in a $15\uppi$ range by successive $\lambda/8$ steps. 
    \textbf{c}, Blue points represent the positions of 
    photonic vortex cores in real space at fixed times, when scanning the phase delay 
    as in panel b. At the times when the photon vortex cores are more distant from the centre (\textit{e.g.}, $t = 3.0$~ps), the phase scan makes their positions rotate with 
    central symmetry, overlapping the fitting circle (black solid line). When the cores are closer to the centre (\textit{e.g.}, $t = 2.6$ and $3.4$~ps), the phase scan makes them describe two ellipses not overlapping anymore with the fitting circle, but composing a 8-shaped line.
    }
\label{FIG_exp_phase}
\end{figure*}

From the photonic point of view, it is possible to perform a fine control of the vortex trajectories upon tuning of the optical phase delay $\varphi_{AB}$ between the pulses $A$ and $B$ (which is directly controlled by the time-delay $t_{AB}$ between the two pulses). Figure~\ref{FIG_exp_phase}a shows the emitted photon density at a fixed time of the 
dynamics ($t = 3.7$~ps), but for varying 
$\varphi_{AB}$ spaced by equal intervals of $\uppi/4$ (corresponding to $\lambda/8$ steps in the physical delay line for the second pulse). Both vortices describe the same circle (black/white line) when sweeping the phase delay, exchanging their positions when changing $\varphi_{AB}$ along a $\lambda/2$ length. Although the distribution of the polariton fluid and the positions of the vortex cores look invariant through such a change, nevertheless, the two vortices can be identified by the continuity of their positions. The ``topological bottle'' surface in the $xyt$ space described by the double vortex strings is shown in Fig.~\ref{FIG_exp_phase}b, in the time range of one and a half Rabi-oscillations period ($t = 2.5 - 3.7$~ps). 
Here the solid spheres represent the position of the two phase singularities, tracked at time intervals of $t = 0.1$~ps ($t = 0.05$~ps in the last part) and sweeping $\varphi_{AB}$ in $\uppi/4$ 
steps. The red and blue solid tubes are 
the specific vortex lines associated to a fixed phase delay (that of the previous figure, sampled with $\delta t = 0.02$~ps), climbing on the surface. The surface apparently resembles a self-twisting double cylinder, but in reality, two different topologies are observed, 
%
%
that can be ascribed to the
anisotropy factor given by the initial splitting of the two vortices, when compared to the cores displacement induced by the pulse $B$. In fact, there is a nodal string of the topological bottle (intersection with a fixed $t$ plane), where the double concentric cylinder undergoes a metamorphosis
into two non-concentric and separated quasi-cylinders: the projection of such $xy\varphi_{AB}$ line onto the $xy$-plane is a 8-shaped line. 
The two different situations are represented in Fig.~\ref{FIG_exp_phase}c, where the circular symmetry is present at given times (\textit{e.g.}, $t = 3.0$~ps), while the 8-shape is visible at $T_{\rm R}/2$ time interval 
from the former (\textit{e.g.}, at $t = 2.6$~ps and $3.4$~ps). These two situations therefore correspond to the times of the Rabi cycle when the twin cores are further from and closer to the centre, respectively.\\

\begin{figure}[htbp]
  \centering \includegraphics[width=1.00\linewidth]{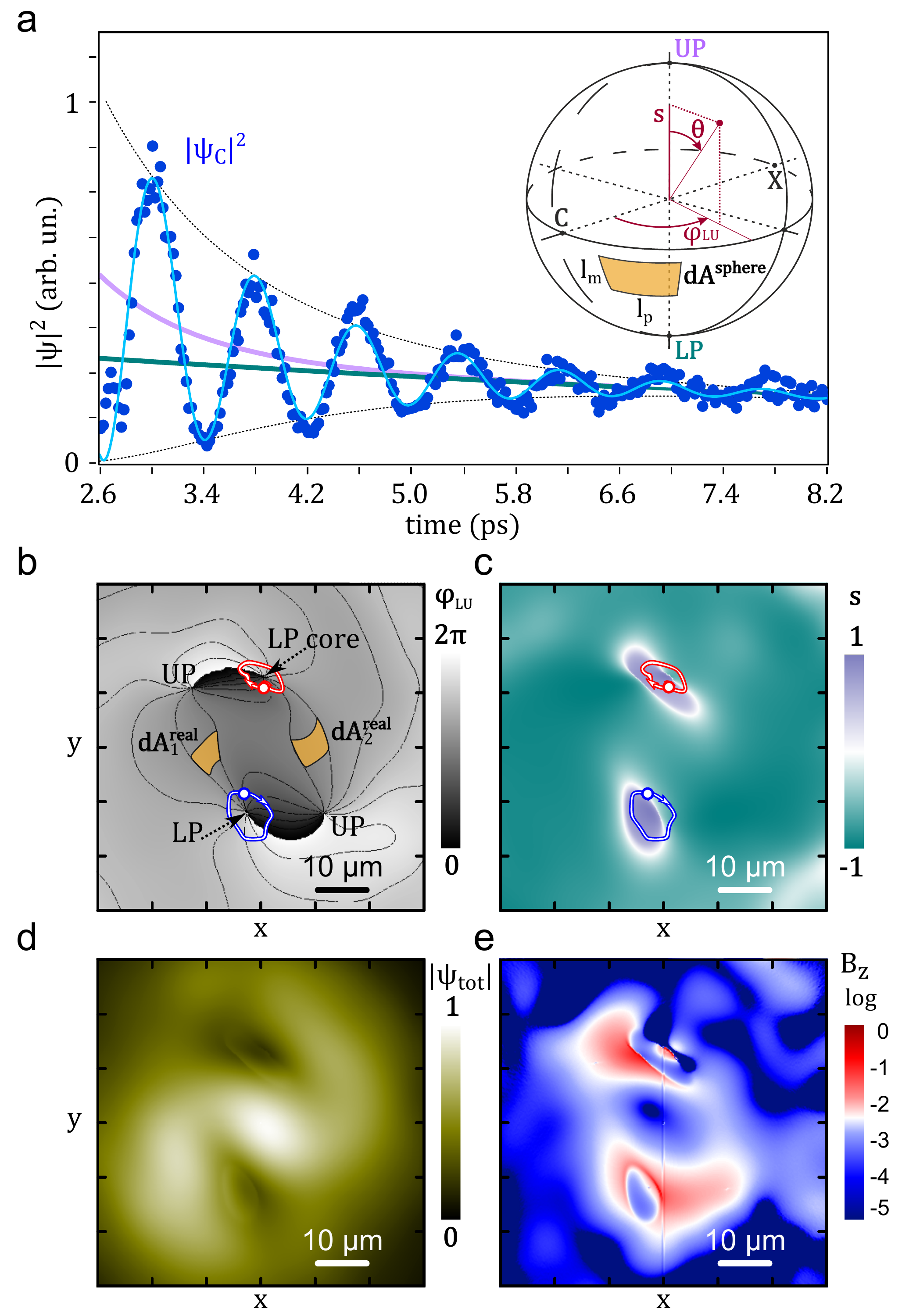}
  \caption{
  \textbf{Intensity oscillations fit and the Bloch metric maps.} 
    \textbf{a}, Photonic oscillations at a specific point (experimental data) are fitted by the two interfering LGs model (see Methods). The procedure allows to retrieve the intensities of the two normal modes in time (solid purple and green lines for the UP and LP modes, respectively) and their relative phase at each point and moment of time. Repeating the procedure all along the spatial domain allows to plot their 2D profiles. Inset: the polariton Bloch sphere with indicated UP, LP states at the poles, photon (C) and exciton (X) states at the equator, and the sphere metric consisting in the azimuthal angle $\varphi_{\rm LU}$ (relative phase) and $s=\cos{\theta}$ (local UP-LP content imbalance).
    An example of area item $\text{d}A^{\rm sphere} = l_{\rm p} l_{\rm m} = \sin{\theta} \text{d}\varphi \text{d}\theta$ is marked by the orange patch.
    \textbf{b,c}, Relative phase map $\varphi_{\rm LU}(x,y)$ and local content imbalance $s(x,y)$ (at $t = 2.7$~ps). The superimposed lines (red, blue) represent the experimental orbits of photonic vortex cores along the second Rabi cycle after the pulse B arrival (time range $t = 2.7 - 3.5$~ps). One clearly observes their rotation around the two relative phase singularities corresponding to the two LP-mode vortex cores. The asymmetry manifests itself both in the fact that the UP mode cores are more distant from each other than the two LP cores, and that the displacement lines of sight 
    are mutually oblique.
    The superimposed orange patches in b mark the real-space area elements $dA^{\rm real}$ corresponding to a given element on the sphere.
    \textbf{d}, Module of the total density $|\Psi_{\rm tot}(x,y)|$ (with $|\Psi_{\rm tot}|^2 = |\psi_{\rm L}|^2 + |\psi_{\rm U}|^2)$. The shape is a footprint mainly of the LP mode.
    \textbf{e}, Normalized Berry curvature map in log scale, retrieved from the maps in panels b,c.
   }
\label{FIG_L2_fit_maps}
\end{figure}

\noindent \textbf{Retrieval of the Bloch sphere metric in real space
}\\
In order to model and discuss the observed spiraling, we first retrieve the topological texture, or quantum states metric, underlying the structured vortex dynamics. Figure~\ref{FIG_L2_fit_maps}a shows that the time-oscillations of the photonic density at a given spatial point can be fitted by the coupled oscillators model (see Methods). 
The two 
normal modes (UP and LP fields) are decaying in time differently, while 
also undergoing a phase shift with different frequencies. 
This evolution can be 
described by introducing the complex frequencies $\omega_{\rm U,L} + {\rm i}\gamma_{\rm U,L}$, so that $\psi_{\rm U,L}(t) = \psi_{\rm U,L}(0)\exp({\rm i}\omega_{\rm U,L}t)\exp(-\gamma_{\rm U,L}t)$. Because of the two different eigenfrequencies, the off-axis displacement of the two vortices 
induced by the arrival of the pulse B appears to different azimuthal positions (since $\varphi_{AB}$ for the UP and LP modes is different for the same time delay $t_{AB}$). The fitting 
of the 
photonic emission $\psi_{\rm C}$ as an interferometric sum of $\psi_{\rm U}$ and $\psi_{\rm L}$ 
allows us to retrieve 
the spatial maps of the relative phase and amplitude 
of the two normal modes (for more details, see Methods section). Since the position of the phase singularities (after the two-pulse excitation) in one of the normal modes is shifted with respect to the other component, the spatial relative phase $\varphi_{\rm LU}(x,y) = \varphi_{\rm U} - \varphi_{\rm L}$ 
profile has the shape of a quadrupole, shown in Fig.~\ref{FIG_L2_fit_maps}b. While the displaced UP and LP vortices are motionless, the two directly observable photon vortex cores appear to be moving in round orbits around the 
singularities of $\varphi_{\rm LU}$, in the specific experimental case 
around those corresponding to the LP vortex cores (see Fig.~\ref{FIG_L2_fit_maps}b). 

To track the position of any quantum state from the polariton Hilbert space (UP, LP, photon, exciton or any other linear combination), one needs to introduce the polariton Bloch sphere (shown in the inset of Fig.~\ref{FIG_L2_fit_maps}a). This sphere has a metric which in turn defines a corresponding metric on the real plane. In real space, we use the local polariton imbalance 
$s(x,y) = (|\psi_{\rm U}|^2-|\psi_{\rm L}|^2)/(|\psi_{\rm U}|^2+|\psi_{\rm L}|^2) = \cos{\theta}$ which, when shown on the Bloch sphere, is the analogue of the $s_3$ Stokes parameter. 
Indeed, $\theta$ represents the polar angle of the corresponding polariton state on the Bloch sphere, while $\varphi_{\rm LU}$ is its azimuthal angle. Thus, in the spatial relative phase profile, the isophase lines $\varphi_{\rm LU}=$~const (thin black lines in Fig.~\ref{FIG_L2_fit_maps}b) correspond to the meridians on the Bloch sphere.  
The north and south poles of the polariton Bloch sphere represent the pure UP and LP eigenstates
(analogously to the right and left circular polarizations on the Poincar\'{e} sphere~\cite{bliokh_geometric_2019}),
whereas the equator represents all possible dynamical states of a pure photon or exciton,
which are continuously transforming into each other at the Rabi frequency $\Omega_{\rm R} = \omega_{\rm U}- \omega_{\rm L}$.
The $s(x,y)$ map at $t = 2.7$~ps is shown in Fig.~\ref{FIG_L2_fit_maps}c. 
The two photon vortex cores, associated to zeros of the photon intensity, are pure excitonic states. 
Furthermore, such zeros
can be understood in terms of a destructive interference between the UP and LP modes, and hence they have to move along the orbit corresponding to $s(x,y) = 0$. 
The orbits obtained from the fitting of the experimental photon intensity (white loops in panel c) fail to precisely retrieve 
the observed photon vortex cores trajectories (red and blue lines), 
due to the high sensitivity of the fit in the weak density areas close to the LP/UP cores,
but the agreement is qualitatively good. While the photonic cores move along the orbits due to the continuous dynamical drift of the relative phase in time $\varphi_{\rm LU} = \varphi_{\rm LU}^0 + \Omega_{\rm R}t$,
the orbits themselves shrink due to the differential decay between the normal modes $\gamma_{\rm R} = \gamma_{\rm U} - \gamma_{\rm L}$. 
From the maps of $s(x,y)$ and $\varphi_{\rm LU}(x,y)$ one can see that each state appearing as one point on the Bloch sphere is mapped twice to the real plane.\\ 

\noindent \textbf{Berry curvature link between the two spaces}\\
We now show that the real-space density of the relative phase and local imbalance isolines, that comprises the metric of quantum states on the Bloch sphere, is described by the Berry curvature---the imaginary part of the quantum geometric tensor~\cite{Berry_1989}---that can be defined for any given target space. 
In condensed matter, the role of momentum-space and real-space Berry curvatures of magnetization 
(or in general, of other vector textures)
was recently highlighted in the anomalous and topological Hall effect (AHE and THE), respectively~\cite{verma_unified_2022,volovik_topological_2019}, affecting the motion of an electron wavepacket.
For the 2D target spaces, such as the one we are dealing with here ($xy$-plane), the generally tensorial Berry curvature becomes a vector with only one non-zero component $B_z$.
In microcavity polaritons, such a curvature has been recently used to express the change of polarization pseudospin  
(\textit{i.e.}, polarization of only one of the polariton modes), thus mapping the density of the Poincar\'{e} coordinates in the reciprocal space~\cite{bleu_measuring_2018,polimeno_tuning_2021}. That was related to the AHE of an accelerating wavepacket~\cite{gianfrate_measurement_2020}.
Here instead, we use 
the Berry curvature  
to express the change of the polariton pseudospin 
(\textit{i.e.}, composition of the full wavefunction in terms of both the polariton modes),
mapping the density of the Bloch sphere coordinates to real space:
\begin{equation}
B_z =\frac{1}{2}\sin\theta(\partial_{x}\theta\partial_{y}\varphi_{\rm LU}-\partial_{y}\theta\partial_{x}\varphi_{\rm LU}).
\label{berryeq}
\end{equation}
Such an expression can be shown to be equivalent to:
\begin{equation}
B_z = \frac{1}{2} \textbf{S}\cdot (\partial_x\textbf{S} \times \partial_y\textbf{S}),
\label{berryeq_vector}
\end{equation}
where $\textbf{S}$ is the pseudospin or Bloch vector with components $(\sin{\theta}\cos{\varphi_{\rm LU}},\sin{\theta}\sin{\varphi_{\rm LU}},\cos{\theta})$ on the sphere. 
The functional form in Eq.~(\ref{berryeq_vector}) was brought into attention
in the Mermin-Ho relations \cite{mermin_circulation_1976}
and was used to express the topological charge density,
or real-space Berry curvature, of different order parameter vectors~\cite{volovik_topological_2019,everschor-sitte_real-space_2014,gobel_beyond_2021,parmee_optical_2022,verma_unified_2022}.
Since Eq.~(\ref{berryeq}) also embeds a cross product of two gradients in real space, $B_z = \text{\textonehalf} \sin\theta ~|\bm{\nabla}\theta \times \bm{\nabla}\varphi_{\rm LU}|$, 
it is straightforward to see (at least in the case of a conformal mapping between the metrics) that the Berry curvature describes how much area of the Bloch sphere is covered when a given area element is spanned in real space.
In other terms, the local density of the sphere surface itself on the real plane can be written as $\text{d}A^{\rm sphere}/\text{d}A^{\rm real} = 2B_z$.
We note that in the present case, each area element on the sphere has a multiplicity $C=2$ in real space (see the color filled areas in both panels Fig.~\ref{FIG_L2_fit_maps}a,b).
The Berry curvature therefore plays the role of linking the full-wavefunction density $|\Psi_{\rm tot}|^2 = |\psi_{\rm L}|^2 + |\psi_{\rm U}|^2$ between the two spaces. In both cases, the meaning of the full wavefunction is to quantify the total number of particles in a small area element: $|\Psi_{\rm tot}|^2 = \text{d}N/\text{d}A$. 
Equating the number of particles in a given real-space area to the corresponding Bloch sphere surface area, 
$dN = C|\Psi^{\rm real}_{\rm tot}(x,y)|^2 \text{d}A^{\rm real} =
|\Psi^{\rm sphere}_{\rm tot}(\varphi,\theta)|^2 \text{d}A^{\rm sphere}$, 
one gets $|\Psi_{\rm tot}^{\rm sphere}|^2 = C|\Psi_{\rm tot}^{\rm real}|^2 / (2B_z)$.
In this derivation, we have assumed a perfect symmetry of the real-space distributions which is not observed in the experiments, but present in the model (see below).
For the 
objects considered in the experiments 
(\textit{i.e.}, superpositions of LG$_{00}$ and LG$_{02}$ beams), the total density has a central maximum and two dips, 
see Fig.~\ref{FIG_L2_fit_maps}d, being a footprint of the UP and LP states at short and long times, respectively, and rotated at intermediate times.
The Berry curvature assumes the shape of a double peak 
(centered in the intensity minima) 
on top of a less intense surrounding ring, with a neat central hole 
corresponding to a saddle point of the isophase lines, as shown in the experimentally-reconstructed Fig.~\ref{FIG_L2_fit_maps}e. 
Recalling the previous connection of the total densities between the two spaces, it is therefore expected that the most populated state on the Bloch sphere is the one for which the pseudospin is associated with the central point of the imprinted LGs in real space (at any given time).

The integral of the Berry curvature over the surface limited by a closed loop in real space represents a solid angle on the sphere. 
The integral of $B_z$ over the whole plane defines the topological charge, the number of times the texture wraps over the sphere, 
also known as the 
Chern number~\cite{xiao} or skyrmion number~\cite{everschor-sitte_real-space_2014,parmee_optical_2022}:
\begin{equation}
C =\frac{1}{4\uppi} \int \textbf{S}\cdot (\partial_x\textbf{S} \times \partial_y\textbf{S}) \text{d}x\text{d}y = \frac{1}{4\uppi} \int 2B_z \text{d}x\text{d}y.
\label{inteq}
\end{equation}
Remarkable examples of Chern numbers can be found for a number of textures in condensed matter.
In the case of superfluid $^3$He-A \cite{mermin_circulation_1976,volovik_topological_2019},
the integrand in Eq.~(\ref{inteq}) 
represents the topological density of the texture of the unit vector $\bm{\hat{\textbf{l}}}$,
that is the relative orbital angular momentum of Cooper pairs in the chiral superfluid. In that case, 
the Mermin–Ho relations establish that
the local vorticity of the continuous vortex is represented by the texture of such a unit vector.
In other terms, they link the circulation of the superfluid velocity 
to the solid angle (\textit{i.e.}, area on the sphere) that is contained inside the real-space loop.
The same can be applied to the current in ferromagnetic BECs where 
the integrand instead refers to the density of magnetization texture in real space.
In those cases also the pseudospins represent physical directions in real space.
In our case, the pseudospin refers to a direction on the auxiliary Bloch sphere of the polariton state.
We have verified that for both the experimental and the model cases described below, 
the entire space integral of the Berry curvature is 4$\uppi$, 
associated to the twofold mapping of the sphere, and remains such at any time despite the reshaping of the mapping. 
This is indeed consistent with the integral of Eq.~(\ref{inteq})
over the whole parameter space giving a net number of two wraps on the sphere. 
In the case of stereographic mapping from the sphere 
to the real plane~\cite{beckley_full_2010,dominici_full-bloch_2021} (\textit{i.e.}, for the single full Bloch beam),
such a wrapping is unitary.
In the current realization, instead, the twofold mapping and the associated Chern number 
are reflected in the term double full Bloch beam
(or, equivalently, dynamical double skyrmion) that we use for such states.
\\

\noindent \textbf{LGs model of the dynamical double full Bloch beam}\\
The essence of the dynamics can be reproduced by means of a model which starts from the overlapping of the two initial LG beams in the normal modes and lets them evolve due to the differential decay and the Rabi oscillations.
Theoretically, the quadrupole 
relative-phase profile formed of four poles (after the second pulse arrival) can be that of a square, a rectangle, a rhombus or a parallelogram, depending on the time-delay $t_{AB}$ which defines the difference in azimuthal displacement of the UP and LP vortex cores.

Experimentally, there are two effects to be taken into account. One is that the initial $\text{LG}_{02}$ pulse has always a small $\text{LG}_{00}$ component (due to imperfect tuning of the $q$-plate or other residual intensity) and the first pulse then imprints in the polariton fluid two vortices which are not ideally overlapped and centered, but have a slight preliminary offset in opposite directions with respect to the center of the beam. Then, the arrival of the second pulse induces a further displacement which may happen along the same or different direction, and this displacement can also differ for the LP and UP vortex cores due to the difference in $\varphi_{AB}$ for the two modes. A second factor is that upon arrival of the second pulse, the differential decay of the modes will have altered 
the amplitudes ratio between the two eigenmodes with respect to the one imprinted by the pulse $A$. 
The lesser content is that of the UP (as this mode decays faster), which makes this mode more sensitive to the displacement induced by the second pulse. As a matter of fact, it can be seen in Fig.~\ref{FIG_L2_fit_maps}b that the UP cores are more offset than the LP ones with respect to the center of the spot. 
The minimal set to account for the initial situation after the two pulses 
is the four-$\text{LG}$ model, one $\text{LG}_{02}$ and one $\text{LG}_{00}$ in both of the normal modes, with the same center and widths $\sigma$ but with
different amplitudes and phases (see Methods). 
The initial displacement of the 
vortex cores inside each of the normal modes is then symmetric, but the direction and amount of such displacement is different between the modes. 

\begin{figure*}[htbp]
  \centering \includegraphics[width=1\linewidth]{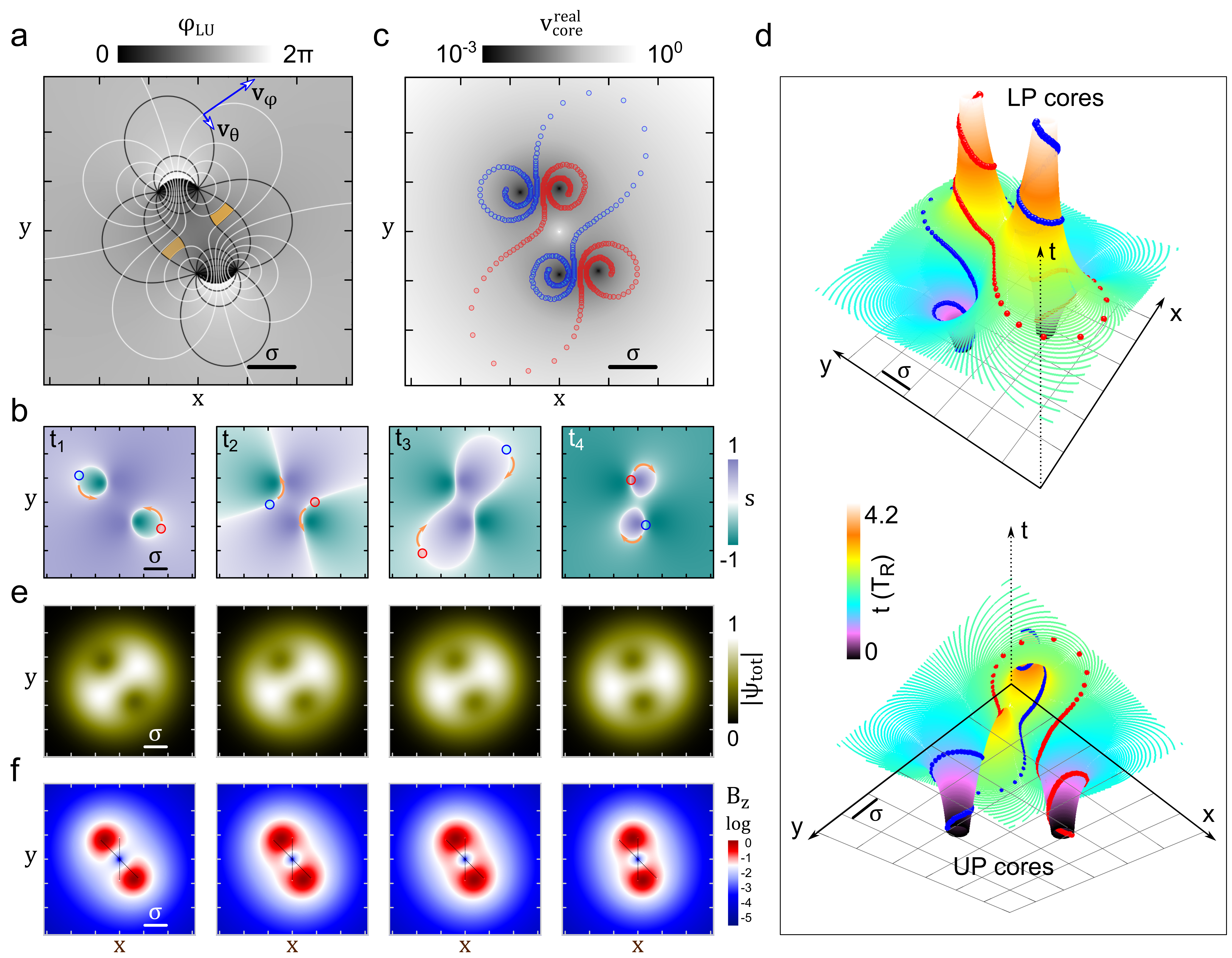}
  \caption{
  \textbf{Model of a spiraling double vortex.}
    \textbf{a}, Spatial map of the relative phase $\varphi_{\rm LU}$ when starting with a larger UP content. The black lines are phase isolines ($\varphi_{\rm LU}$ every 30$^\circ$), the white lines are isolines of the local imbalance ($s$ from 0.8 to -0.8 with step 0.2). The superimposed orange patches are an example of real-space area elements corresponding to a given sphere area element. The two arrows (not is scale) depict the velocity components in the motion of a given pseudospin.
\textbf{b}, Spatial maps of the local UP-LP imbalance $s(x,y)$ at four different times, with its reshaping due to the differential decay. 
The instantaneous orbits of the photon vortex cores (red and blue circles) follow the white contour lines $s = 0$. 
Two initial orbits expand till they join at infinity, then become an eight-shaped line until they finally recoil into two separated petals again.
\textbf{c}, Spatial map of the amplitude of velocity of the moving vortex cores (as well as of any other pseudospin state). 
The trajectory of the vortex cores in the photonic field is tracked (blue and red open circles) at regular time intervals.
\textbf{d}, Perspective views of the isocontent $s = 0$ line evolving in time, drawing a 3D surface with two valleys and two hills. The early (bottom) cones open out from the position of the UP vortex cores, the later (top) cones wrap into the position of the LP cores. The overlapped 
red and blue circles mark the photon vortex cores trajectory. 
\textbf{e}, Absolute value of the 
full wavefunction $|\Psi_{\rm tot}(x,y)|=\sqrt{|\psi_{\rm L}|^2 + |\psi_{\rm U}|^2}$ at the moments of time corresponding to panel b. 
\textbf{f}, Normalized Berry curvature map in log scale for the same times as in panels b,e. The two dashed black lines mark the UP and LP vortex pairs line of sight. 
The scale bar is equal to the width of the beams $\sigma = 16~\upmu \text{m}$ in all the panels, the Rabi period $T_{\rm R} =0.785$~ps.
See also Methods and the Supplementary Movie 2.}
\label{FIG_L2_models}
\end{figure*}
A typical (corresponding to experiment) scenario is reproduced in
Fig.~\ref{FIG_L2_models}a where the more asymmetric condition of the four poles makes a parallelogram. The vortex quadrupole is shown together with the relative phase (black) and local imbalance (white) isolines.
It is worth to note the electric-like (conservative) shape of the isophase lines and the magnetic-like (solenoidal) shape of the isocontent lines.
Interestingly, even while the isolines $s(x,y) =$~const are not circles, they are still perpendicular to the relative phase isolines. 
We express this by using the condition on the gradients of the Bloch sphere coordinates in real space, 
$\bm{\nabla}\theta~\perp~\bm{\nabla}\varphi_{\rm LU}$.
In other terms, the Bloch sphere's parallels and meridians that are by definition orthogonal on the sphere surface, remain mutually orthogonal in real space, and likewise any other angle relation is preserved, meaning that the mapping between the Bloch sphere of polariton states and the real space is fully conformal. However, it cannot be described simply by a stereographic projection or a homeomorphism as in a fundamental 
full Poincar\'{e}\cite{beckley_full_2010} or single full Bloch beam case~\cite{dominici_full-bloch_2021}, since the one-to-one link is lost (as already mentioned, each pseudospin is mapped twice to the real space). 

The theoretically modeled evolution of the UP-LP local imbalance map $s(x,y)$ is shown in Fig.~\ref{FIG_L2_models}b for four selected moments in time. The initially closed orbits ($s = 0$, white loop regions) expand until when, approximately at the moment of the global populations equality, they become two diagonal edge lines connecting to each other at an infinite distance~\cite{molina-terriza_vortex_2001}, after which they start to shrink again, forming a single 8-shaped line. Finally, they separate in the central point leaving two closed orbits again. The experimentally-retrieved panel Fig.~\ref{FIG_L2_fit_maps}c taken at $t=2.7$~ps corresponds to the final panel in Fig.~\ref{FIG_L2_models}b. 
In essence, the motion of the cores (as well as of any other pseudospin state) has two orthogonal components, and its velocity can be expressed as 
$\textbf{v}_{\rm ps}^{\rm real} = \textbf{v}_{\theta} + \textbf{v}_{\varphi}$, 
where $\textbf{v}_{\theta} = (\gamma_{\rm R}\sin{\theta}/|\bm{\nabla}{\theta}|)  \bm{\hat{\nabla}}{\theta}$
and
$\textbf{v}_{\varphi} = (\Omega_{\rm R}/|\bm{\nabla}{\varphi_{\rm LU}}|)  \bm{\hat{\nabla}}{\varphi}_{\rm LU}$
(with $\bm{\hat{\nabla}}{\theta} = \bm{\nabla}{\theta} / |\bm{\nabla}{\theta}|$ and
$\bm{\hat{\nabla}}{\varphi}_{\rm LU} = \bm{\nabla}{\varphi}_{\rm LU} / |\bm{\nabla}{\varphi}_{\rm LU}|$
the unit vectors along the direction of the two gradients, 
see also Methods and the blue arrows in Fig.~\ref{FIG_L2_models}a for their schematic representation, not in scale).
Such two components are conservative and solenoidal, respectively.
The velocity and the trajectory of a given pseudospin state are further characterized as following a constant angle $\alpha$ with respect to the imbalance isolines, defined as $\tan{\alpha} = \gamma_{\rm R}/\Omega_{\rm R}$. 
This is derived from the pseudospin of a fixed point in real space evolving on the Bloch sphere along loxodromes, the curves forming a constant angle with 
its parallels, and from the full conformal mapping.
The amplitude of the velocity in real space is shown in a log scale map in Fig.~\ref{FIG_L2_models}c. The four darker spots represent the four poles of larger gradients, where the photon vortex cores move slower (in the same panel we superimposed also the vortices cores trajectories). 
The moving cores can end up their spiral in one of the two final orbits, depending on which side they are immediately before the orbits' separation (which is set by the initial conditions and $\gamma_{\rm R}/\Omega_{\rm R}$ ratio), or eventually cross in the very center.
This crossing possibility is also expressed by the central point being a saddle point~\cite{berry_geometry_2001}, which means that the relative phase is defined, but its isolines have a singularity of their direction. Due to the conformality of the mapping, the same is valid for the local imbalance isolines. A full view of the orbit line evolution in time is presented in Fig.~\ref{FIG_L2_models}d, depicted as two different perspectives for the sake of clarity, with the central point visible as a 3D saddle point also of the $s=0$ surface 
in the $xyt$ domain. 
The steeper slope of the surface at earlier and later times correspond to the smaller velocity of the swirling photon 
vortices cores close to the LP and UP cores positions.

The time dynamics of the dips in the total density 
and the peaks of the Berry curvature 
consists in 
drifting from the initial positions on top of the two UP vortex cores to the positions of the two LP cores, as shown in Fig.~\ref{FIG_L2_models}e,f, respectively. This 
reshaping is only due to the differential decay, since the Rabi rotation is
homogeneously swapping the sphere meridians, without altering the metric density in real space. 
In more symmetric cases, the Berry curvature can as well assume the shape of a perfect ring, 
due to orthogonal displacements between the UP and LP vortex pairs (achievable upon proper phase delay). 
The two maxima can also become diverging curvature points when the global population imbalance is mainly in one of the two modes, \textit{i.e.}, when initializing the system at the UP mode energy, or at long times, when the differential decay leads to the mainly LP content. Such points can also be thought of as the limit of the projected sphere contracting to a zero radius, representing two monopoles. In other terms, this is equivalent to the continuous vortices becoming two singular vortices (as the system loses its spinorial degree of freedom and becomes a one-component wavefunction).
As anticipated, 
using the space integral of the Berry curvature in Eq.~(\ref{inteq}) 
gives a skyrmion number equal to two, which remains the same 
during the Rabi spinning and reshaping in time, 
indicating the topological conservation of the textures.\\

\noindent \textbf{Speed of the  photon vortex cores in real space}\\
As a very interesting property of dynamical full Bloch beams,
we derived also the speed of the polariton pseudospins in real space, 
revealing their link to velocities of the polariton states spiraling on the Bloch sphere,
again via the Berry curvature connection of the metrics (see Methods). 
Such a relation can be written as 
\begin{equation}\label{vel_Bz}
v_{\rm ps}^{\rm real} =
\frac{v_{\rm ps}^{\rm sphere}}{\sqrt{2B_z}} =
\frac{\sin\theta}{\sqrt{2B_z}}
\sqrt{\gamma_{\rm R}^2+ \Omega_{\rm R}^2}.
\end{equation}
Here, the spatial map of the velocities remains constant in time despite $B_z(x,y)$ is reshaping, compensated by the simultaneous reshaping of the polar angle $\theta$, whereas the second multiplier in (\ref{vel_Bz}) can be viewed as the absolute value of the complex-valued 
frequency $\Tilde{\Omega}_{\rm R} = 
\Omega_{\rm R} +i\gamma_{\rm R}$.

The connection between real-space Berry curvature of a topological texture 
and some physical properties has been known for different systems.
In superfluid $^3\text{He-A}$,
 the celebrated Mermin-Ho relations link the vorticity (curl) of the superfluid velocity $\textbf{v}_{\rm s}$
to the texture density of the unit vector $\bm{\hat{\textbf{l}}}$~\cite{mermin_circulation_1976}
(or of the magnetization in ferromagnetic multicomponent BECs~\cite{yukawa_su_2023}).
For simplicity, in the case of a 2D $xy$-dependence of the order parameter, that is, a vorticity parallel to $z$
(as also in the case of 2D polariton fluids), these relations assume the following form~\cite{parts_phase_1995}
\begin{equation}\label{mermin_one}
\bm{\nabla} \times \textbf{v}_{\rm s} \propto
\bm{\hat{\textbf{l}}}\cdot (\partial_x\bm{\hat{\textbf{l}}} \times \partial_y\bm{\hat{\textbf{l}}}),
\end{equation}
where the right-hand side is the $B_z$ associated with $\bm{\hat{\textbf{l}}}$. 
We now point out one main difference between the two expressions.
In Eq.~(\ref{mermin_one}), the curl of the total superfluid velocity is directly proportional to the real-space Berry curvature of the unit vector of the order parameter. In the case of Eq.~(\ref{vel_Bz}), the velocity of the pseudospin states (not its curl) is, instead, inversely proportional to the square root of the real-space Berry curvature of the polariton Bloch sphere. 
Here it is useful to recall that in the case of a homogenenous $\bm{\hat{l}}$ texture, or of a single-component fluid or condensate, the superfluid velocity is proportional to the gradient of the phase of the wavefunction. In the language of optics or exciton-polaritons, it is therefore equivalent to the group velocity ($|\textbf{v}_{\rm g}| = \partial{\omega}/\partial{k} \propto k = |\bm{\nabla}\varphi|$, in the parabolic approximation for one component). 
In these cases, such a velocity is irrotational and its curl is zero everywhere apart from the diverging point of a singular vortex core (where, conversely, the density needs to be zero). In two-component wavefunctions, the concept of phase needs to be modified, leading to the notion of total superfluid velocity~\cite{yukawa_su_2023,volovik_topological_2019}, 
and the vorticity may be diffused (\textit{i.e.}, a finite curl of velocity in an extended area where the total density doesn't need to be zero). This is the case of $v_s$ appearing in Eq.~(\ref{mermin_one}).
However, in our case and the situation described by Eq.~(\ref{vel_Bz}), we are 
not looking at the total phase. 
The associated total group velocity plays a role, as it is known, in the motion 
of the total density, whose reshaping is further affected in two-component cases
by the relative decay term, or differential radiative losses between the two components.
While the total density is still plotted in our representation, we are not focusing on its driving terms.
Here, we are rather dealing with the velocity of the pseudospin states themselves moving on top of the total density. To this end, since the motion of the pseudospin can be thought of in terms of interference between the normal modes, their speed is more analogue to a phase velocity ($|\textbf{v}_{\rm ph}| = \omega/k \propto k^{-1} = |\bm{\nabla}\varphi|^{-1}$). 
This intuitively explains that the velocity of the pseudospins 
is inversely proportional to the Berry curvature (if one thinks of the latter as the gradient of a 2D-valued phase).
Another difference is that the overall velocity in Eq.~(\ref{vel_Bz}), as mentioned before, has both the conservative and the solenoidal terms linked to the Berry curvature. Furthemore, when projecting the full wavefunction onto one specific component (such as the emitted photon field), one can study the motion of the (singular) vortex cores themselves inside that component (rather than looking at the group velocities, or fluid velocities, around the cores, for a fixed vortex configuration, be it singular or continuous).

At the same time, the velocity described in Eq.~(\ref{vel_Bz}) 
is also different from the terms involved in the AHE and THE, that represent velocity and force terms that are directly proportional to the momentum- and real-space Berry curvatures, respectively~\cite{verma_unified_2022}. In the AHE, 
an accelerated wavepacket aquires a transverse velocity term too, proportional to 
the momentum-space Berry curvature (of magnetization vector in condensed matter, and of polarization for polaritons~\cite{gianfrate_measurement_2020}).
On the other hand, the real-space and mixed spatio-temporal Berry curvatures of the magnetization vector result in the emergent magnetic and electric fields, respectively~\cite{volovik_linear_1987}.
In the former case, also known as THE, an electron wavepacket acquires an extra force proportional to the real-space skyrmion texture~\cite{everschor-sitte_real-space_2014}.
In the latter case, known as spinmotive force (and requiring a time derivative of the relevant vector, \textit{e.g.}, local magnetization), it is the moving skyrmion that generates a force perpendicular to its moving direction~\cite{matsuki_thermoelectric_2023}.
Here we are not dealing with velocities obtained by (or forces exerted on) external objects, but with the self-evolution of the pseudospin texture in a two-component wavefunction (externally generated and then driven by an effective Rabi field, as indicated below).

It is now interesting to focus on the velocity of the observable photon vortex cores, which represent pure excitonic states that lie on the equator of the Bloch sphere, \textit{i.e.}, 
$\sin\theta=1$. 
In this case, Eq.~(\ref{vel_Bz}) reduces to 
\begin{equation}\label{vel_core}
v_{\rm core}^{\rm real} =
\frac{|\Tilde{\Omega}_{\rm R}|}{\sqrt{2B_z}},
\end{equation}
where the vortices cores' velocity only depends on the complex Rabi frequency and is inversely proportional to the (square root of the) local and instantaneous Berry curvature, changing during the motion.
It can also be seen that there is no mass or any other prefactor in the right-hand side of Eq.~(\ref{vel_core}) (while this was omitted in the case of Eq.~(\ref{mermin_one})).
This is a general relation that is assumed to hold also in the case of superposition of two optical LG (or even differently shaped) beams with different frequencies (when looking at the dynamics on a plane orthogonal to the propagation direction).\\


\noindent \textbf{CONCLUSIONS}\\
When dealing with a two-component fluid, its full wavefunction can be expressed in terms of a total density and total phase, plus a further pseudospin unit vector or 2D valued phase. Such a pseudospin 
is representable on a Poincar\'{e} or Bloch sphere, and can be mapped by the sphere texture (or metric) to the real space. 
In a fundamental unitary skyrmion (or single full Bloch beam~\cite{dominici_full-bloch_2021} and, analogously, Poincar\'{e} beam~\cite{beckley_full_2010}) 
the one-to-one mapping is possible via a stereographic projection, 
a conformal homeomorphism between the sphere metric and real space, 
while in more complex cases even the conformal feature is not maintained~\cite{dominici_shaping_2021}. 
Our experiments implement for the first time dynamical pseudospin textures going beyond the unitary skyrmion case in a polariton fluid, based on multiple-vortex configurations. Such textures are set into a spinning and reshaping dynamics thanks to the polariton Rabi oscillations 
and the non-Hermitian feature imparted by the decay. 
The Rabi frequency as an operator, similarly to other Hamiltonians, 
can be thought of as an effective magnetic field, 
as for example was proposed in the optical Maxwell-Schr\"{o}dinger formalism~\cite{kuratsuji_maxwell-schrodinger_1998,singh_synthetic_2023} for the dynamics of the polarization pseudospin on the Poincar\'{e} sphere.
For the polariton state, the Rabi field is along the vertical axis $\bm{\hat{s}}_3$ of the Bloch sphere,
resulting in the precession of the polariton pseudospin around such axis (\textit{i.e.}, along a parallel)
\begin{equation}\label{effective_Rabi}
\frac{d\textbf{S}}{dt} = \Omega_{\rm R} \bm{\hat{s}}_3 \times \textbf{S} + (\gamma_{\rm R} \bm{\hat{s}}_3 \times \textbf{S}) \times \textbf{S},
\end{equation}
while the second term adds the decay-induced drift along a vertical plane 
(\textit{i.e.}, along a meridian). 
The two terms result in the solenoidal and conservative velocities of the pseudospin states in real space, respectively, as a feature directly inherited from the associated parallels and meridians on the sphere (and mediated by the conformal mapping).
The form of Eq.~\ref{effective_Rabi} is equivalent to the Landau-Lifshitz (LL)~\cite{landau_theory_1992,prigogine_statics_2007} equation that describes the Larmor precession of the magnetic moment of an object (or the magnetization vector) around the external (or effective) magnetic field (that is, on the other hand, also analogue to the precession of a tilted mechanical gyroscope in an external, torque-exerting gravitational field).
From a physical point of view, the LL equation has a term that corresponds to the torque exerted by the effective field on the magnetic moment forcing it to precession, and a second damping (relaxation) term that represents another torque that pushes the magnetization in the direction of the field.
Also in the case of $^3\text{He-A}$, there are equivalent and even more complex equations~\cite{vollhardt_superfluid_1990}, describing how 
the vector $\bm{\hat{\textbf{l}}}$ varies in time due to the external rotation of the system, the magnetic field in case of locked spin-orbital configurations, and other torques due to incoherent fermionic excitations.
Here, the two main terms are represented by the Rabi frequency (responsible of a cyclic conversion of photons into excitons and so on) and the differential decays of polaritons (responsible for the unidirectional evolution from the upper polariton to the lower polariton mode).
Furthermore, in the LL case, both the magnetic moment and the magnetic field are directions in real space.
Here, the effective Rabi field as well as the decay term, 
are directions on the pseudospin sphere. 
The field and decay are assumed to be uniform, and the pseudospin texture varying in real space
(\textit{e.g.}, creating a unitary or multiply-charged skyrmion).
It is this effective field and associated decay that rules the dynamics of the pseudospin and reshapes the texture.
This results 
in the observable photonic vortex cores (being associated to a specific state on the equator of the sphere) following a peculiar spiraling motion. Such an effective field preserves the initial conformal feature of the texture mapping at all times, despite being non-stereographic and non-homeomorphic. Conversely, it is also possible to show that, since the composing beams only change in intensity and phase, the transformation maintains the initial ingredients of the LGs superposition. We empirically devise the conformal feature as a general property of concentric $\text{LG}$ beams of the same width bearing cowinding charges (of different absolute value). 

Furthermore, we apply for the first time the Berry curvature concept to the polariton Bloch-sphere mapping to real space, specifically describing it for dynamical double skyrmions.
The Berry curvature here represents the density of the Bloch-sphere metric 
(the variation of the pseudospin) in the target space (real space). 
As such, its integral over the whole parameter space 
represents the number of times the texture wraps around the Bloch sphere, in our case equal to two (in both experiments and model cases) and conserved during time,
despite the spinning and reshaping of the texture and of the curvature itself.
One could think of the initial state in the model as composed by two skyrmions centered in the positions of the UP mode vortex cores. The decay strains the two initial skyrmions into four merons and then transforms them back again into two skyrmions, centered in the LP vortex cores at later times, while the Rabi oscillations transform them from Bloch- to Neel-type continuously. This is consistent with the topological charge equal to two at all times.
It would be possible to use counter-polarized optical pulses to achieve a four-component fluid and build a one-to-one mapping of the Bloch and Poincar\'{e} spheres, realizing a dynamical full Poincar\'{e} beam. Each polarization would be present at a given time in a different position, and any point of space would see a sequence of polarizations in time, too. In another configuration, having a different (asymmetric content between the two pulses) superposition (such as for the simplest full Bloch beam possible, a LG$_{01}$ in the UP mode and a LG$_{00}$ in the LP mode), would lead to the observation of an initial vortex in the photonic component (emission), spiraling out of the polariton cloud (and vice versa, spiraling in, when using the opposite scheme). Similar reasonings pertain to any generic LGs superposition of two (or more) optical beams with different (complex) frequencies. These structured beams could be used to drive THE and spinmotive forces when applied to optically anisotropic particles.
Experimentally, it is interesting to see how a pseudospin texture can manifest itself into an observable point-like object, here the photonic vortex cores. Their motion depends on how the pseudospin trajectories are linked between the two spaces. Along the sphere, these directions are loxodromes with an angle directly linked to the ratio of the two components of the complex Rabi frequency. On the other hand, we show that the same velocity is projected onto the real plane with a speed defined by the Berry curvature. 
Intuitively, the pseudospin gradient in real space can be further thought of as a pseudomomentum, similarly to the gradient of a standard phase, which is now a 2D-valued phase. One of the meanings of the Berry curvature is hence to describe the amplitude of the pseudomomentum vector. The real-space velocity of the pseudospins' motion is inversely linked to such a pseudomomentum (similar to standard phase velocity being inversely proportional to the momentum in one-component fields).
This allows us to derive a 
compact equation for the real-space velocity of the observable photonic vortex cores, which ultimately depends only on the complex Rabi frequency and is inversely proportional to the square root of the local Berry curvature.


\vspace{0.2cm}

\begin{small}

\noindent {\normalsize \textbf{METHODS}}\\
\noindent \textbf{Experimental methods.}
The polariton device used here is fabricated by means of MBE technique and consists in an AlGaAs $2\lambda$ MC with three In$_{0.04}$Ga$_{0.96}$As quantum wells of 8 nm placed at the antinodes of the cavity mode field. The cavity is placed inside two distributed Bragg reflectors made of 21 and 24 pairs of alternated $\lambda/4$ AlAs and GaAs layers. The polariton modes are at 836.2 nm and 833.2 nm (LP and UP mode, respectively), their splitting is 3~nm (5.4~meV) at zero momentum which converts into a Rabi period of $T_{\rm R} =0.780$~ps. The lifetime of the normal modes is $\tau_{\rm L}=1/\gamma_{\rm L}\sim 10$~ps and $\tau_{\rm U}=1/\gamma_{\rm U}\sim 2$~ps for the lower and upper modes, respectively. The experiments are performed in a region of the sample clean from defects. The device is kept at a temperature of 10 K inside a closed-loop He cryostat.
The resonant beam is a 130~fs pulse laser with 80~MHz repetition rate and a 8~nm bandwidth. The central energy is tuned at approximately 835 nm in order to overlap both the LP and UP branch energies. The photonic beam is passed through a $q$-plate device~\cite{marrucci_optical_2006,karimi_efficient_2009} with winding charge 2, in order to obtain a double optical vortex which is sent onto the microcavity sample at normal incidence. 
Once imprinted, the polariton vortices just decay and oscillate in intensity in the photonic component keeping a fixed position. The second pulse is sent by means of a small Michelson-Morley (MM) scheme interferometer in the excitation side, with controllable time delay. The second pulse is derived from the unconverted portion of the beam crossing the $q$-plate, through polarization filtering.
The $q$-plate is a patterned Liquid Crystal (LC) device that is driven by a oscillating voltage in the kHz range. 
The applied voltage controls the degree of LC tilting along the vertical direction, hence the anisotropy degree along the patterned directions and ultimately the $q$-plate conversion efficiency. An impinging $\text{LG}_{00}$ beam with right or left circular polarization is totally or partially converted into the opposite circular polarization with a wavefront shaped into an $\text{LG}_{02}$ mode. So there can be a tunable amount of residual original $\text{LG}_{00}$ beam. Upon using a different polarization filtering scheme in the small MM interferometer arms, such a residual portion can be both left in the pulse A and used for pulse B.
In summary, pulse A will consist of a $\text{LG}_{02}$ and $\text{LG}_{00}$ beams, whose relative amplitudes control the initial two cores displacement (the same in the two normal modes), while pulse B will only consist of an $\text{LG}_{00}$ pulse used to further and 
differentially split the two cores (\textit{i.e.}, along different directions in the normal modes).
An ultrafast time resolved detection scheme is based on the off-axis digital holography, where the emission from the sample plane is focused on an imaging camera together with a reference beam. The reference pulse is derived by the resonant laser beam in a Mach-Zehnder scheme interferometer scheme and is not focused but passed through a small iris, so that it becomes a homogeneous and flat front when arriving on the camera. The digital fast Fourier transform (FFT) filtering is applied to the interferograms and this procedure allows to reconstruct the amplitude and phase of the emitted photonic wavefunction. The FFT filtering can be applied directly in the laboratory in order to aid for the setting of the measurements.
Further details on the polariton sample or digital holography technique can be found in previous works~\cite{dominici_ultrafast_2014,colas_polarization_2015,dominici_full-bloch_2021,Satyajit_geometric_2019} and descriptions therein.\\

\noindent \textbf{Nonlinear regime effects.}
We checked the changes added in the experiments when deviating from the linear regime, in this and previous works configurations and add here a brief note on those. When increasing the power we effectively see two main effects. First, a differential shift of the oscillations frequency between spatial regions of low and small density, that we ascribe to the nonlinear energy increase of the exciton line, mainly reflected in the LP mode blueshift. On the other hand, the UP mode energy acts for this sample mainly in a self-compensating way, due to the exciton blueshift and the partial loss of coupling. Second, increasing further the power we observe the damping of the oscillations amplitude, due to an increased linewidth of the UP mode, associated to its faster decay and/or dephasing. This damping of the oscillations converts, in the current vortex experiments, into a faster damping of the swirling cores orbits.\\


\noindent \textbf{Interference-fitting procedure for two coupled oscillators.}
In the linear regime, it is possible to express the evolution of the observable photon density at each point in space as a mean value plus a modulation term, \textit{i.e.}, in terms of interference between the LP and UP eigenmodes:
\begin{equation*}\label{interference}
|\psi_{\rm C}|^2 = \frac{|\psi_{\rm L}|^2 + |\psi_{\rm U}|^2}{2} + |\psi_{\rm L}||\psi_{\rm U}|\cos(\varphi_{\rm LU}^0+\Omega_{\rm R} t).
\end{equation*}
The numerator defines also the total density, expressed as $|\Psi_{\rm tot}|^2 = |\psi_{\rm L}|^2 + |\psi_{\rm U}|^2$.
The UP and LP modes decay are included upon adding $|\psi_{\rm U,L}(t)| = |\psi_{\rm U,L}(0)|\exp(-\gamma_{\rm U,L}t)$. 
Imposing fixed values for
$\gamma_{\rm U}$, $\gamma_{\rm L}$, and $\Omega_{\rm R}$, and treating $|\psi_{\rm L}({\bf r},0)|$, $|\psi_{\rm U}({\bf r},0)|$, and $\varphi_{\rm LU}^0({\bf r})$ as fitting parameters, we can fit the time traces at each point $(x,y)$ as in Fig.~\ref{FIG_L2_fit_maps}a starting from a time taken after the setting sequence of both pulses $A$ and $B$, and retrieve the profiles of the UP and LP polariton fields (which are not directly observable), as well as their relative phase spatial profile shown in Fig.~\ref{FIG_L2_fit_maps}b. 
The fitting is performed along seven Rabi periods (from 2.6 ps to 8.2 ps). The experimental maps in Fig.~\ref{FIG_L2_fit_maps}c-e are retrieved by using the fitted values in the expressions for the local imbalance parameter $s$, amplitude of the total density $|\Psi_{\rm tot}|$ and Berry curvature $B_z$, respectively.\\

\noindent \textbf{Interfering LGs model.}
The initial condition and its evolution 
are reproduced by use of the four $\text{LGs}$ packets model,
representing the vortex and Gaussian beam imprinted in the upper and lower polariton fields. Assuming for simplicity a zero detuning between the cavity photon to exciton mode, the bare modes are hence obtained as linear combinations of the normal modes
$\psi_{\rm C,X} = (\psi_{\rm U} \pm \psi_{\rm L})/\sqrt{2}$.
We write the spatial profiles of the fields in complex polar coordinates $(r,\phi)$, where $r=|x+{\rm i}y|$ and $\phi = \arg(x+{\rm i}y)$. The initial state can be written as a combination of the four LGs,\\

$
\psi_{\rm U,L}(0) = A_{\rm U,L}\text{LG}_{02} + B_{\rm U,L}\text{LG}_{00}. \\
$

\noindent Expanding the LGs forms and adding the time evolution on the right side, it can be written\\

$
\psi_{\rm U,L} =  {\rm e}^{-r^2/2\sigma^2} \left[ 
 A_{\rm U,L} \frac{r}{\sigma} {\rm e}^{{\rm i}2\phi }
 + B_{\rm U,L} {\rm e}^{{\rm i} \varphi_{AB}^{\rm U,L}} 
 \right] 
 {\rm e}^{({\rm i}\omega_{\rm U,L}-\gamma_{\rm U,L})t}.\\
$

\noindent We have used the same width $\sigma$ and same center in the origin for all the four components. The four coefficients (real positive) represent the amplitude of the UP and LP modes set by the two pulses $A$ and $B$. 
Despite the two laser pulses have the same frequencies content, we use $B_{\rm U}/B_{\rm L} > A_{\rm U}/A_{\rm L}$. In fact, at the time arrival of second pulse $B$ (about 2 ps time delay), the UP/LP content set by pulse $A$ is already decreased due to the differential decay. The pulse delay also sets a phase delay between the plain Gaussian and the vortex packet that is in general different in each of the two modes, $\varphi_{AB}^{\rm U} \neq \varphi_{AB}^{\rm L}$. These phase delays control the direction of the displacement of the vortex cores in each respective (UP or LP) mode.
Upon changing time delay on a few wavelengths scale, the relative displacement directions of the UP and LP remain the same, meaning they are rotated of the same amount. For such reason we use in the previous sections the simplified notation $\varphi_{AB}$.
This simple model gives the same results than would be derived with a full coupled Schr\"{o}dinger equations (cSE) model, apart from neglecting the dispersion and diffraction effects and the transients from the incoming photon pulses being injected into the polariton components. We are not interested in the sub-ps transient and the beams are wide enough in order to neglect any dispersive/diffraction effects in the 10 ps time scale range. 
The cSE model was used in order to verify the behavior of the Rabi-oscillating vortices against the LGs model in previous configurations~\cite{dominici_full-bloch_2021,dominici_shaping_2021}. The parameter values used in the model are: $A_{\rm U} = 6$, $A_{\rm L} = 2$, $B_{\rm U} = 10$, $B_{\rm L} = 2$, $\varphi_{AB}^{\rm U} = \uppi/2$, $\varphi_{AB}^{\rm L} = 0$ and $\sigma = 16~\upmu \text{m}$.
The used relative frequency and decay are: 
$\Omega_{\rm R} = \omega_{\rm U} - \omega_{\rm L} = 8~\text{ps}^{-1}$ and $\gamma_{\rm R} = \gamma_{\rm U} - \gamma_{\rm L} = 0.8~\text{ps}^{-1}$, resulting in a ratio $\gamma_{\rm R}/\Omega_{\rm R} = 0.1$. The time frames shown in Fig.~\ref{FIG_L2_models} are $t_{1,2,3,4} = 0.89, 1.38, 1.58~\text{and}~ 2.11~\text{ps}$, corresponding to a global content imbalance $S_{1,2,3,4} = 0.5$, $0.16$, $0.0$ and $-0.4$, respectively. The global imbalance is defined as $S=\frac{\int |\psi_{\rm U}|^2d{\bf r}-\int|\psi_{\rm L}|^2d{\bf r}}{\int|\psi_{\rm U}|^2d{\bf r}+\int|\psi_{\rm L}|^2d{\bf r}}$.\\

\noindent \textbf{Pseudospin evolution on the sphere.}
We have defined the pseudospin using the polar and azimuthal angles $\theta, \varphi$, that define the latitude and longitude on the Bloch sphere of polaritons, respectively.
The polar angle can also be written as a Stokes-like parameter, 
the local imbalance $s = \frac{|\psi_{\rm U}|^2-|\psi_{\rm L}|^2}{|\psi_{\rm U}|^2+|\psi_{\rm L}|^2} = \cos{\theta}$ 
while the azimuthal angle is represented by the relative phase $\varphi_{\rm LU}=\varphi_{\rm U}-\varphi_{\rm L}$.
A point in real space has a given pseudospin at time $t_0$. Such state will change due to the complex relative (\textit{i.e.}, Rabi) frequency. The azimuthal angle of the pseudospin on the sphere will change according to $\varphi_{\rm LU}(t) = \varphi_{\rm LU}(0) + \Omega_{\rm R} t$. Its arc line speed along a parallel is hence $\text{d}l_{\rm p}/\text{d}t = \Omega_{\rm R}\sin{\theta}$. The polar angle will change due to the differential decay term. Combining the expression of the two decays, $\frac{|\psi_{\rm U}|}{|\psi_{\rm L}|}(t) = \frac{|\psi_{\rm U}|}{|\psi_{\rm L}|}(0) e^{-\gamma_{\rm R}t}$, that can be written in terms of the Stokes parameter $s(t) = \tanh\{-\gamma_{\rm R}t -\text{atanh}[s(0)]\}$. 
The initial $s(0)$ is arbitrary, here it depends both on the initialization conditions and on the specific point in space.
The speed of variation of the parameter is $\text{d}s/\text{d}t = \frac{-\gamma_{\rm R}}{\text{cosh}^2\{-\gamma_{\rm R}t -\text{atanh}[s(0)]\}} = -\gamma_{\rm R}(1-s^2) = -\gamma_{\rm R} \sin^2{\theta}$ (where we used the hyperbolic derivative and the $\text{cosh}^2 = \frac{1}{1-\text{tanh}^2}$ relation). Recalling that $\text{d}s = -\sin{\theta} \text{d} \theta$, the arc length speed along a meridian is $\text{d}l_{\rm m}/\text{d}t = \text{d}\theta/\text{d}t = \frac{1}{-\sin{\theta}} \text{d}s/\text{d}t = \gamma_{\rm R} \sin{\theta}$. The pseudospin evolves on the sphere moving at an angle $\alpha$ with respect to the parallels such that $\tan{\alpha} = \frac{\text{d}l_{\rm m}/\text{d}t}{\text{d}l_{\rm p}/\text{d}t} = \gamma_{\rm R}/\Omega_{\rm R}$. The angle is hence constant, describing a so called loxodrome (we used this in a previous work~\cite{dominici_full-bloch_2021}, without showing the current demonstration). The total velocity of the polariton state evolution on the sphere can be expressed as $v_{\rm ps}^{s\rm phere} = \text{d}l/\text{d}t = \sin{\theta}\sqrt{\gamma_{\rm R}^2 + \Omega_{\rm R}^2}$. The term in the squared root can be thought as the module of a complex Rabi frequency $\Tilde{\Omega}_{\rm R} = \Omega_{\rm R} +i\gamma_{\rm R}$.
The evolution holds for the pseudospin state in any point of real space, regardless of the specific mapping of the Bloch sphere to real space.\\

\noindent \textbf{Conformal mapping.} 
We have verified that the superposition of LGs with the same width, center and sign of all the winding charges among the two modes always convert into a conformal mapping of the pseudospin sphere to real space (regardless of the packets' total number, amplitudes, relative phases and absolute winding charges). 
Naturally, linearly dependent combinations of coefficients between the two modes result in a 0D mapping on the sphere, that is the same pseudospin on the whole space.
Conformal mapping means that parallels and meridians of the Bloch sphere are mapped to two families of curves mutually orthogonal everywhere on the real plane. Since we didn't find a complete reference treating this property, we point to more specialized bibliography or to a future work. In our multifrequency LGs models we neglect diffusive and dispersion effects, the composing LGs only change their relative amplitudes and phase, hence the conformal feature remains at all times. This feature can be expressed in terms of the real-space gradients of the two pseudospin angles, 
$\bm{\nabla}{\theta}~\perp~\bm{\nabla}{\varphi}_{\rm LU}$. 
By application of the chain rule, we know that the motion of a pseudospin in real space due to, \textit{e.g.}, the solely Rabi effect happens along the direction of the gradient of $\varphi_{\rm LU}$ with speed $v_{\varphi} = \frac{\partial d}{\partial t} = \frac{\partial d}{\partial \varphi} \cdot \frac{\partial \varphi}{\partial t} = (\frac{\partial \varphi}{\partial d})^{-1} \cdot \frac{\partial \varphi}{\partial t} = \Omega_{\rm R}/|\bm{\nabla}\varphi_{\rm LU}|$
(where we used $d$ for the distance along such a direction, and the compact notation $\varphi = \varphi_{\rm LU}$ for simplicity).
Applying the same on the other component, and from the evolution speed of the polar angle, $\frac{\partial \theta}{\partial t} = \gamma_{\rm R} \sin{\theta}$, we end up with the total vector velocity
$\bf{v}_{\rm ps}^{\rm real} = \bf{v}_{\theta} + \bf{v}_{\varphi}$ being $\bf{v}_{\theta} = (\gamma_{\rm R}\sin{\theta}/|\bm{\nabla}{\theta}|) \bm{\hat{\nabla}}{\theta}$
and
$\bf{v}_{\varphi} = (\Omega_{\rm R}/|\bm{\nabla}\varphi_{\rm LU}|) \bm{\hat{\nabla}}{\varphi}_{\rm LU}$.
The second term does not change in time, because the relative phase only changes by a homogeneous dynamical term due to the Rabi frequency, so its gradient is constant. The first term does not change as well, because our specific mapping is fully conformal, such that $|\bm{\nabla}{\theta}| =  \sin{\theta}|\bm{\nabla}\varphi_{\rm LU}|$, so that $v_{\theta}/v_{\varphi} = \gamma_{\rm R}/\Omega_{\rm R}$, so the trajectory of a given pseudospin is a loxodrome in real space too (with respect to the $\varphi_{\rm LU},\theta$ curvilinear coordinates).  
Looking at the expression of the Berry curvature in Eq.~(\ref{berryeq}),
it contains the vector product of the two pseudospin angles gradients in real space, $B_z = \text{\textonehalf} \sin\theta ~|\bm{\nabla}{\theta} \times \bm{\nabla}{\varphi}_{\rm LU}|$. 
Such a product represents the sphere areal density on real space. 
Since the two gradients are here orthogonal, it is also $B_z = \text{\textonehalf} \sin\theta ~|\bm{\nabla}{\theta}| |\bm{\nabla}{\varphi}_{\rm LU}| = \text{\textonehalf}~|\bm{\nabla}{\theta}|^2 = \text{\textonehalf} \sin^2\theta ~|\bm{\nabla}{\varphi}_{\rm LU}|^2$. The two velocities can then be written as
$v_{\theta} = \frac{1}{\sqrt{2B_z}} \gamma_{\rm R} \sin{\theta}$ and
$v_{\varphi} = \frac{1}{\sqrt{2B_z}} \Omega_{\rm R} \sin{\theta}$,
while for the module of the total velocity
$v_{\rm ps}^{\rm real} = \frac{1}{\sqrt{2B_z}}\sin{\theta}
\sqrt{\gamma_{\rm R}^2+ \Omega_{\rm R}^2}$.
It is possible to see that the Berry curvature is linking the velocities between the two spaces (real plane and Bloch sphere), recalling that $v_{\rm ps}^{\rm sphere} = \sin{\theta}\sqrt{\gamma_{\rm R}^2 + \Omega_{\rm R}^2}$.
Hence, the speed of the spatial motion of a pseudospin and of the evolution on the sphere of a spatial point's pseudospin are linked by $v_{\rm ps}^{\rm real} = \frac{1}{\sqrt{2B_z}} v_{\rm ps}^{\rm sphere}$.
On the other hand, since the total density represents the number of total particles in a given area, the Berry curvature also plays a role in linking the total density between the two spaces, $|\Psi_{\rm tot}^{\rm sphere}|^2 = \frac{C}{2B_z} |\Psi_{\rm tot}^{\rm real}|^2$, where $C$ is the topological multiplicity factor discussed in the main text.\\

\noindent \textbf{Total phase and continuity equation.} 
For the sake of clarity and completeness, we derive the continuity equation for the total density of particles $n_{\rm tot} = n_{\rm U} + n_{\rm L}$, where $n_{\rm U} =|\psi_{\rm U}|^2$ and $n_{\rm L} =|\psi_{\rm L}|^2$ are the local densities of particles in the two components.
With this notation the local imbalance reads $s = \frac{n_{\rm U} - n_{\rm L}}{n_{\rm U}+n_{\rm L}} = \frac{n_{\rm U} - n_{\rm L}}{n_{\rm tot}}$.
It is also possible to consider the total phase 
$\varphi_{\rm tot} = \varphi_{\rm U}+\varphi_{\rm L}$ 
in parallel with the relative phase $\varphi_{\rm LU} = \varphi_{\rm U}-\varphi_{\rm L}$. 
For convenience, we also introduce the total decay rate 
$\gamma_{\rm tot} = \gamma_{\rm U}+\gamma_{\rm L}$ 
in addition to the already discussed relative decay $\gamma_{\rm R} = \gamma_{\rm U} - \gamma_{\rm L}$, where $2\gamma_{\rm U}$ and $2\gamma_{\rm L}$ are the particles decay rates in each component. 
The continuity equation can be written as 
$\text{d}n_{\rm tot}/\text{d}t = - \bm{\nabla}\cdot\textbf{j}_{\rm U} - 2\gamma_{\rm U} n_{\rm U}
- \bm{\nabla}\cdot\textbf{j}_{\rm L} - 2\gamma_{\rm L} n_{\rm L}$, where
$\textbf{j}_{\rm U,L} = n_{\rm U,L} \textbf{v}_{\rm U,L}$ are the density flow in each component 
and $\textbf{v}_{\rm U,L} = \frac{\hbar}{m_{\rm U,L}} \bm{\nabla}\varphi_{\rm U,L}$ the superfluid velocities, associated to the respective effective masses $m_{\rm U,L}$ and phase gradients $\bm{\nabla}\varphi_{\rm U,L}$.
The total flow of particles can be written in terms of the total density and velocity, $\textbf{j}_{\rm tot} = \textbf{j}_{\rm U} + \textbf{j}_{\rm L} = n_{\rm tot} \textbf{v}_{\rm tot}$, from which the total superfluid velocity is defined as
$\textbf{v}_{\rm tot} = \frac{n_{\rm U}\textbf{v}_{\rm U} + n_{\rm L}\textbf{v}_{\rm L}}{n_{\rm U} + n_{\rm L}}$.
The continuity equation can then be written in terms of the total and relative quantities, rather than with respect to the single components, 
\begin{equation*}\label{continuity}
{\text{d}n_{\rm tot}}/{\text{d}t} = 
-\bm{\nabla} \cdot (n_{\rm tot}\textbf{v}_{\rm tot}) - (\gamma_{\rm tot} + s \gamma_{\rm R}) n_{\rm tot} .
\end{equation*}
Also the total superfluid velocity contained in such expression can be written in terms of the total and relative quantities, 
\begin{equation*}\label{tot_velo}
\textbf{v}_{\rm tot} =
\frac{\hbar}{2m} (\bm{\nabla}\varphi_{\rm tot} + s \bm{\nabla}\varphi_{\rm LU}) ,
\end{equation*}
where we used the approximation of equal effective masses $m = m_{\rm U} = m_{\rm L}$ (that is not strictly true for polaritons but valid at small momenta and for small photon-exciton detuning). It is hence possible to see that the total phase and its gradient is in general involved in the motion and reshaping of the total density profile, similarly to what happens in one-component cases. However, the total density is in principle also affected by the content imbalance, if there is a difference between the two superfluid velocities (\textit{i.e.}, if $\textbf{v}_{\rm U} \neq \textbf{v}_{\rm L}$). Finally, while in one-component cases, a uniform decay rate is only affecting the absolute density, not its shape, in two-component cases the total density is also reshaping due to the different decay rates of the components. The dynamics of the total density in Fig.~\ref{FIG_L2_models}e is in fact only due to this effect 
(while the model neglects the density redistribution due to the group velocities, 
that is a valid approximation, as said, for the used beams width and time scales).
Apart from the role on the total density, the total superfluid velocity satisfies the Mermin-Ho relation in Eq.~(\ref{mermin_one}), upon replacing the 
$\bm{\hat{\textbf{l}}}$ vector with the polariton pseudospin $\textbf{S}$.
It is possible to note that this is due to the second term in the $\textbf{v}_{\rm tot}$ expression, that has a nonzero curl, thanks to the $s$ factor that appears before the gradient of the relative phase, while for the first term, its curl is zero. 
The circulation of the total velocity has hence a discrete part in $2\uppi$ multiples, depending on the 
total-phase singularities enclosed by the path, and a continuous part that depends on the relative phase, whose gradient is weighted by the $s$ local imbalance.\\

\noindent \textbf{Data Availability:} 
All data needed to evaluate conclusions are present in the paper. 
Original datasets and additional data related to this paper may be requested from the corresponding author on reasonable request. Supplementary Movies 1 and 2 contain the full dynamics of experimental and model double full Bloch beams.
\noindent \textbf{Code Availability}: The code used to analyze data or implement the model 
may be requested from the corresponding author on reasonable request.




\end{small}



%


\begin{small}
\noindent \textbf{Acknowledgements:} 
We acknowledge Romuald Houdr\'{e} and Alberto Bramati for the microcavity sample, Lorenzo Marrucci and Bruno Piccirillo for the $q$-plate devices, Paolo Cazzato for technical support, Charlie Leblanc, Miguel Alonso, Giovanni I.~Martone, Grigory E.~Volovik 
and sir Michael V.~Berry for interesting comments. 
\noindent \textbf{Funding:} We thank for funding 
the Italian Ministry of University (MUR) PRIN project “Interacting Photons in Polariton Circuits”– INPhoPOL (grant 2017P9FJBS); 
the PNRR MUR project “National Quantum Science and Technology Institute” - NQSTI (PE0000023); 
the PNRR MUR project “Integrated Infrastructure Initiative in Photonic and Quantum Sciences” - I-PHOQS (IR0000016);
the Russian Foundation for Basic Research (CNR-RFBR Joint Bilateral Agreement Triennal Program 2021–2023 and CNR-RFBR Joint Project No. 20–52–7816).
N.V. acknowledges the financial support of MEPhI Priority 2030 Program.
A.R. acknowledges support from National Science Center, Poland (PL), Grant No. 2016/22/E/ST3/00045. 
\noindent \textbf{Author Contributions:} L.D. proposed the effect and realized the experiments, with the help of M.D.G., D.B., D.S. and G.G. in realizing the set-up; L.D. analysed the data, proposed the model and implemented numerical simulations, in coordination with N.V., A.R. and D.C.; L.D., A.R., N.V and F.P.L. discussed the theoretical aspects; L.D. and N.V. drafted the manuscript, with the supervision of F.P.L., and all the authors discussed the results and the final manuscript editing.
\noindent \textbf{Competing interests:} 
No competing interests.\\

\end{small}



\onecolumngrid

\renewcommand{\figurename}{\textbf{Supplementary Movie }}
\renewcommand\thefigure{\textbf{S\arabic{figure}}}
\setcounter{figure}{0}


\noindent {\textbf{SUPPLEMENTARY MATERIALS.}}\\

\begin{figure}[h]
  \centering \includegraphics[width=0.57\linewidth]{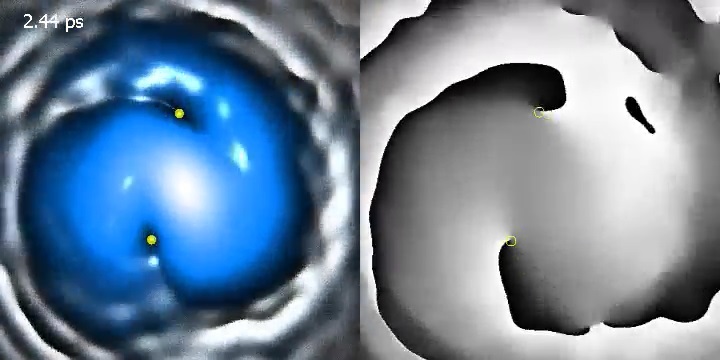}
  \caption{
  Experimental Rabi-oscillating double vortex seen in the photonic component as in Fig.~\ref{FIG_exp_time}. Photonic amplitude and phase in a $100 \times 100~\mu\text{m}^2$ area, with 20 fs time step over a 4.20 ps time span. The second photonic pulse is arriving at around $t = 2.4~\text{ps}$. The vortex cores positions are marked with yellow dots and circles in the amplitude and phase maps.
[\href{https://arxiv.org/src/2202.13210v3/anc/video_S1_L2_exp.mp4}{Video SM1}].
}
\label{snapshot_video_S1}
\end{figure}
\begin{figure}[h]
  \centering \includegraphics[width=0.65\linewidth]{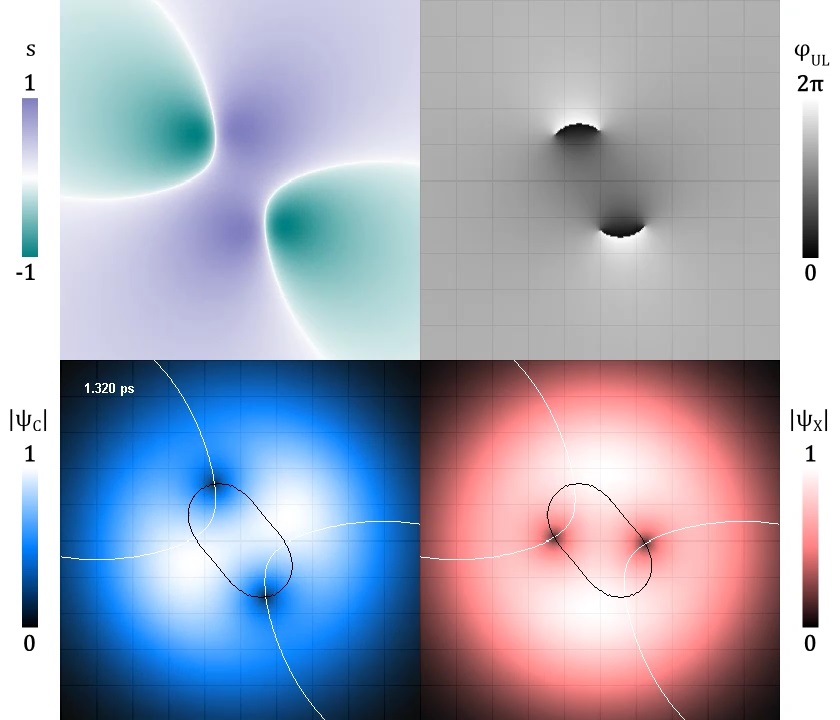}
  \caption{
  The $\text{LG}$s model for the double Rabi-oscillating vortex, comprising the maps and dynamics
of four relevant quantities.
The four panels show the 
local imbalance $s(\bm{r},t)$ (purple-green map) and the relative phase $\varphi_{LU}(\bm{r},t)$ (black and white map)
on the top row,
as in Fig.~\ref{FIG_L2_models}a,b,
together with the photon and exciton amplitudes $|\psi_\mathrm{C,X}(\bm{r},t)|$
(blue and red maps, respectively) in the second row.
The positions of the displaced UP/LP cores are corresponding to the poles of the vortex quadrupole in the relative phase map.
The spiraling motion of the photon and exciton cores, seen as holes in their respective fields, are tracked by the crossing points  
between the $s=0$ isocontent (white)
and the $\varphi_{LU} =0$ and $\varphi_{LU} = \pi$ 
isophase lines (black lines).
[\href{https://arxiv.org/src/2202.13210v3/anc/video_S2_L2_model_maps.mp4}{Video SM2}].
}
\label{snapshot_video_S2}
\end{figure}
\begin{figure}[h]
  \centering \includegraphics[width=0.55\linewidth]{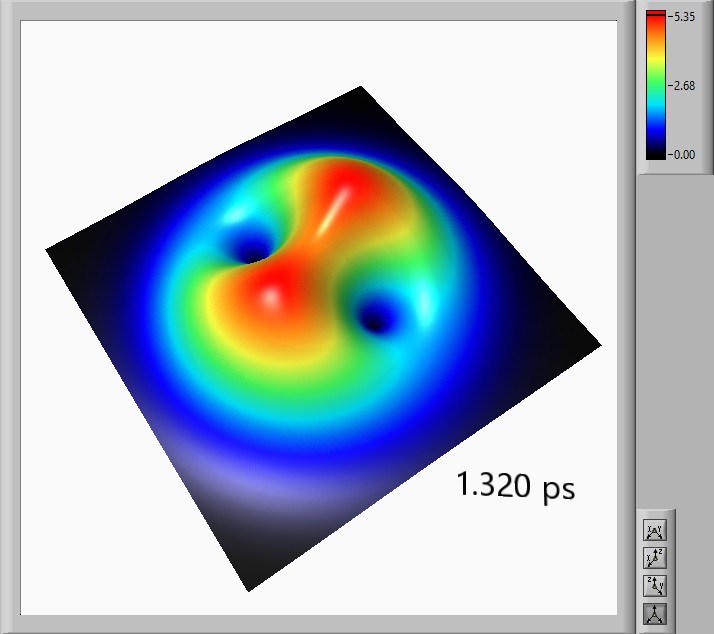}
  \caption{
  Photonic amplitude $|\psi_\mathrm{C}(\bm{r},t)|$ seen as a 3D image.
The amplitude is plotted in an arbitrary false colour bar on the $z$ axis,
in the same $xy$ and time domain as the previous model case.
[\href{https://arxiv.org/src/2202.13210v3/anc/video_S3_3D_photonic.mp4}{Video SM3}].
}
\label{snapshot_video_S3}
\end{figure}
%







\end{document}